\documentclass[ejs, preprint]{imsart}
\usepackage{graphicx}
\usepackage{caption}
\usepackage{bbm}
\usepackage{subcaption}
\usepackage{amsmath}
\usepackage{amssymb}
\usepackage{color}
\usepackage{bm}
\usepackage{float}
\usepackage{multicol}
\usepackage[table]{xcolor}
\usepackage{enumitem}
\usepackage{tabularx}
\usepackage{amsthm}
\usepackage{multirow}
\usepackage{algorithm}
\usepackage{algorithmic}
\usepackage{threeparttable}
\usepackage{natbib}
\bibpunct{(}{)}{;}{a}{,}{,}
\usepackage{etoolbox}
\usepackage[colorlinks,citecolor=blue,urlcolor=blue]{hyperref}

\newcommand{\defi}{\mathop{=}\limits^{\Delta}}      

\makeatletter

\patchcmd{\NAT@citex}
  {\@citea\NAT@hyper@{%
     \NAT@nmfmt{\NAT@nm}%
     \hyper@natlinkbreak{\NAT@aysep\NAT@spacechar}{\@citeb\@extra@b@citeb}%
     \NAT@date}}
  {\@citea\NAT@nmfmt{\NAT@nm}%
   \NAT@aysep\NAT@spacechar\NAT@hyper@{\NAT@date}}{}{}

\patchcmd{\NAT@citex}
  {\@citea\NAT@hyper@{%
     \NAT@nmfmt{\NAT@nm}%
     \hyper@natlinkbreak{\NAT@spacechar\NAT@@open\if*#1*\else#1\NAT@spacechar\fi}%
       {\@citeb\@extra@b@citeb}%
     \NAT@date}}
  {\@citea\NAT@nmfmt{\NAT@nm}%
   \NAT@spacechar\NAT@@open\if*#1*\else#1\NAT@spacechar\fi\NAT@hyper@{\NAT@date}}
  {}{}

\startlocaldefs
\theoremstyle{theorem}
\newtheorem{lemma}{Lemma}

\theoremstyle{definition}

\newtheorem{remark}{Remark}
\endlocaldefs

\begin{document}

\begin{frontmatter}

\title{Solution Path Clustering with Adaptive Concave Penalty\thanksref{T1}}
\runtitle{Clustering with Concave Penalty}
\thankstext{T1}{Revised and resubmitted to \textit{Electronic Journal of Statistics}.}


\begin{aug}
  \author{\fnms{Yuliya} \snm{Marchetti}\thanksref{t1}\ead[label=e1]{yuliya@stat.ucla.edu}}
  \and
  \author{\fnms{Qing}  \snm{Zhou}\thanksref{t2,t3}\corref{}\ead[label=e2]{zhou@stat.ucla.edu}}
  
   \address{Department of Statistics, University of California, Los Angeles \\
		8125 Math Sciences Bldg.,
		Los Angeles, California 90095 \\
          \printead{e1}; \printead*{e2}}

  \thankstext{t1}{Supported by NSF graduate research fellowship and UCLA dissertation year fellowship.}
  \thankstext{t2}{Supported in part by NSF grants DMS-1055286 and DMS-1308376.}
  \thankstext{t3}{Corresponding author.}

  \runauthor{Marchetti and Zhou}


\end{aug}

\begin{abstract}
Fast accumulation of large amounts of complex data has created a need for more sophisticated statistical methodologies to discover interesting patterns and better extract information from these data. The large scale of the data often results in challenging high-dimensional estimation problems where only a minority of the data shows specific grouping patterns. To address these emerging challenges, we develop a new clustering methodology that introduces the idea of a regularization path into unsupervised learning. A regularization path for a clustering problem is created by varying the degree of sparsity constraint that is imposed on the differences between objects via the minimax concave penalty with adaptive tuning parameters. Instead of providing a single solution represented by a cluster assignment for each object, the method produces a short sequence of solutions that determines not only the cluster assignment but also a corresponding number of clusters for each solution. The optimization of the penalized loss function is carried out through an MM algorithm with block coordinate descent. The advantages of this clustering algorithm compared to other existing methods are as follows: it does not require the input of the number of clusters; it is capable of simultaneously separating irrelevant or noisy observations that show no grouping pattern, which can greatly improve data interpretation; it is a general methodology that can be applied to many clustering problems. We test this method on various simulated datasets and on gene expression data, where it shows better or competitive performance compared against several clustering methods. 
\end{abstract}




\tableofcontents

\end{frontmatter}

\section{Introduction}
\label{introduction}

Cluster analysis allows us to group a collection of objects into subsets such that objects within a subset are similar to each other, while objects in different subsets are dissimilar from each other. Clustering is widely used in exploratory data analysis and has a great variety of applications ranging from biology, astrophysics to social sciences and psychology. Clustering is a first step in knowledge discovery where prior information is rarely available to a researcher. It is also instrumental for visualization of complex data and for partitioning a dataset into more homogeneous groups in which simpler models might be adequate. At the same time, clustering can enable discovery of unknown groups or associations, providing deeper insights into the data. For example, in biological research, clustering can help determine which genes are associated with particular cellular functions or phenotypes and can help isolate subclasses of diseases for targeted treatments.  

There exist a great variety of clustering methods. A majority of them rely on minimizing some loss function, usually by an iterative procedure, like the well known k-means algorithm, while other methods recursively organize objects into trees, like the popular hierarchical clustering. Spectral clustering techniques are based on graph theory and matrix decomposition and are gaining popularity as being simple, accurate, and able to find non-convex clusters \citep{luxburg2007, hastie2009}.
Along with k-means and its variants, there also exist mixture-likelihood clustering approaches \citep{mclachlan2002, fraley2002, yeung2001} that assume an underlying statistical model for the data and maximize the likelihood function with the EM algorithm or MCMC methods \citep{bensmail1997, oh2007}. Yet other methods take modified or combined approaches, for example, self-organizing maps \citep{kohonen1990}, dp-means \citep{kulis2011}, CLICK \citep{sharan2003}, gene shaving \citep{hastie2000}, pclust \citep{wangH2008}, support vector clustering \citep{benhur2001} to name just a few. Clustering literature is vast, and surveys of clustering methodologies are usually specialized. A recent comprehensive review of basic and new clustering methodologies is presented in \citet{aggarwal2013}, and more general developments and trends in clustering are covered in \citet{jain2010}.  

Increasingly large and complex datasets, such as those in gene expression analysis or data mining, have created the need for new efficient approaches to clustering. Data now often contain large amounts of both noisy observations and irrelevant variables. Most existing clustering methods do not address the problems of identifying noise and selecting meaningful variables. Only recently, researchers have shown that simultaneously accommodating for the presence of noisy or irrelevant observations can immensely improve clustering results and provide a better interpretation of the patterns in the data \citep{tseng2005, thalamuthu2006}. A number of new clustering algorithms have been proposed, the most popular of which include the resampling-based tight clustering method \citep{tseng2005}, penalized weighted k-means \citep{tseng2007}, and model-based clustering \citep{fraley2002}. Other methods that take noise into account include adap\_Cluster \citep{desmet2002}, k-clips \citep{maitra2009}, DWCN \citep{shen2010}, trimmed k-means \citep{garcia2008} and other robust clustering algorithms \citep{soltanolkotabi2013, forero2012}. Another challenging task in clustering in general is the specification of the number of clusters, which is required as an input for most of the existing methods. When the number of clusters is a required input, the obtained solution is prone to error, especially when a dataset is large and complex. There are a number of methods that suggest rules for choosing the number of clusters. For a recent overview, please refer to \citet{fang2012}. 

A different class of methodologies that have gained popularity for high-dimensional complex datasets is sparsity regularization techniques. Some examples of such penalization methods include the lasso \citep{tibshirani1996}, the elastic net \citep{zou2005}, the group lasso \citep{yuan2006}, the fused lasso \citep{tibshirani2005}, SCAD \citep{fan2001}, SparseNet \citep{mazumder2011}, and MC+ \citep{zhang2010}. These methods are mostly used in linear and generalized linear models for identifying useful predictors among a large number of covariates. They introduce a penalty to a loss function to find sparse solutions in challenging problems. Efficient optimization methods exist that compute the entire regularization path or a particular solution for a penalized loss function, such as the least angle regression \citep{efron2004} and coordinate descent \citep{friedman2007, friedman2010, wu2008}.

We propose a novel solution path clustering (SPC) method that can be applied to a wide range of data settings, including high dimensionality and the presence of noisy or irrelevant observations, as this method is able to isolate these observations in singleton or very small clusters. Our algorithm minimizes a penalized quadratic loss function under the minimax concave penalty (MCP) \citep{zhang2010}. The regularization allows us to obtain sparse solutions and to construct a solution path with a decreasing number of clusters, which eliminates the need to specify the number of clusters as an input parameter. The method minimizes a non-convex objective function via the majorization-minimization (MM) algorithm \citep{lange2004} coupled with block coordinate descent. We also develop adaptive data-driven strategies for selecting the penalty parameters along a solution path so that the SPC algorithm has in effect only one tuning parameter for initializing the path. Overall, SPC is a simple, easily implemented and relatively fast algorithm that has worked well in practice, although its convergence properties are to be established in future work. 

Penalized estimation has been previously utilized in clustering mostly for variable selection, such as in gene expression analysis \citep{pan2007, wang2008, xie2008, zhou2009, guo2010, witten2010, sun2012}. Unlike SPC, these methods assume a given number of clusters and select useful variables to partition objects. Very recently, a number of authors introduced penalized clustering methods that are similar to our method in that they also impose a penalty on the pairwise differences between cluster centers and can generate solution paths that do not require the specification of the number of clusters. \citet{pelckmans2005}, \citet{hocking2011}, \citet{lindsten2011} and \citet{chi2013} suggested minimizing an objective function where the penalty on the differences between the cluster centers is convex and, thus, the resulting algorithms are guaranteed to converge globally. The potential severe bias in the cluster center estimates from these procedures are handled through penalty weights. The first three referenced papers focus primarily on the optimization of objective functions with convex penalties and do not discuss the choice of the penalty parameter, the detection of noise, performance on high-dimensional data, or the impact of the penalty weights on the clustering results. \citet{chi2013} have improved on these convex clustering methods and have provided a general unified algorithm for solving such problems. They have also noted that the solution paths obtained from convex clustering can be unsatisfactory if the weights are not selected properly.

Finally, \citet{pan2013} have proposed a penalized regression-based clustering method (PRclust) using a novel non-convex penalty on the pairwise differences in order to alleviate the possible bias of convex penalties. The authors have re-parametrized the objective function to ensure the convergence of the coordinate descent algorithm to a stationary point. PRclust, however, has not been shown to handle noise and large high-dimensional datasets. In contrast to our adaptive selection of penalty parameters, \citet{pan2013} mainly focus on determining the number of clusters by searching over a pre-specified grid of three penalty parameters resulting from the re-parametrization, which might not be efficient for large complex datasets.       

The remainder of this paper is organized as follows. Section~\ref{formulation} provides a general formulation of a clustering problem under a concave penalty. In Section~\ref{spc} we develop our clustering algorithm with adaptive solution path construction. In Section~\ref{simstudy} we use simulated data to illustrate and compare the SPC algorithm to several clustering methods. In Section~\ref{solselection} we briefly discuss solution selection for SPC. In Section~\ref{geneexp} we apply SPC to a gene expression dataset from mouse embryonic stem cells to show the performance on bigger real data. Finally, Section~\ref{discussion} contains further discussion and future research directions.

\section{Formulation}
\label{formulation}

Let $Y = (y_{im})_{n \times p}$ be an observed data matrix, where $y_i = (y_{i1}, \ldots, , y_{ip}) \in \mathbb{R}^p$ represents the $i$th object.  Assuming that the underlying model for $y_i$, $i=1, \ldots, n$, is multivariate Gaussian with a mean parameter $\theta_i \in \mathbb{R}^p$ and a constant diagonal covariance matrix $\sigma^2 I_p$, we propose to cluster the $n$ objects into an unknown number of clusters $K$ by minimizing a penalized $\ell_2$ loss function. This is achieved by the use of sparsity regularization on the difference between pairwise mean parameters $d( \theta_i, \theta_j ) = \| \theta_i - \theta_j \|_2$. Our goal is then to minimize
\begin{align}
	\label{eq:genloss}
	& \ell( \theta ) = \sum_{i=1}^n \| y_i - \theta_i \|_2^2 + \lambda \sum_{ i < j } \rho \left( \| \theta_i - \theta_j \|_2 \right),
\end{align}
over $\theta = (\theta_1, \ldots , \theta_n)$, where $\lambda > 0$ and $\rho(\cdot)$ is some penalty function. With a careful choice of $\rho(\cdot)$ we can achieve sparsity such that $\| \hat{\theta}_i - \hat{\theta}_j \|_2$ is arbitrarily small when $\lambda$ is sufficiently large, where $( \hat{\theta}_1, \ldots, , \hat{\theta}_n )$ is the minimizer of \eqref{eq:genloss}. An important advantage of this formulation, for the situations when there is very little prior knowledge about the data, is that the number of clusters $K$ does not need to be specified beforehand. This regularization also makes it possible to naturally separate noisy objects that should not belong to any cluster and to prevent them from erroneously merging into other clusters, as demonstrated by the simulation study results, in particular the adjusted rand index scores in Figures~\ref{ARI_bign}, \ref{ARI_bigp}, and \ref{nonspherical} and the cluster assignment plots in Figures~\ref{fig4} and \ref{nonconvex}. See Section~\ref{simstudy} for more details.

\begin{figure}[t]
	\centering
	\begin{subfigure}[b]{0.45\textwidth}
		\centering
		\includegraphics[width=\textwidth]{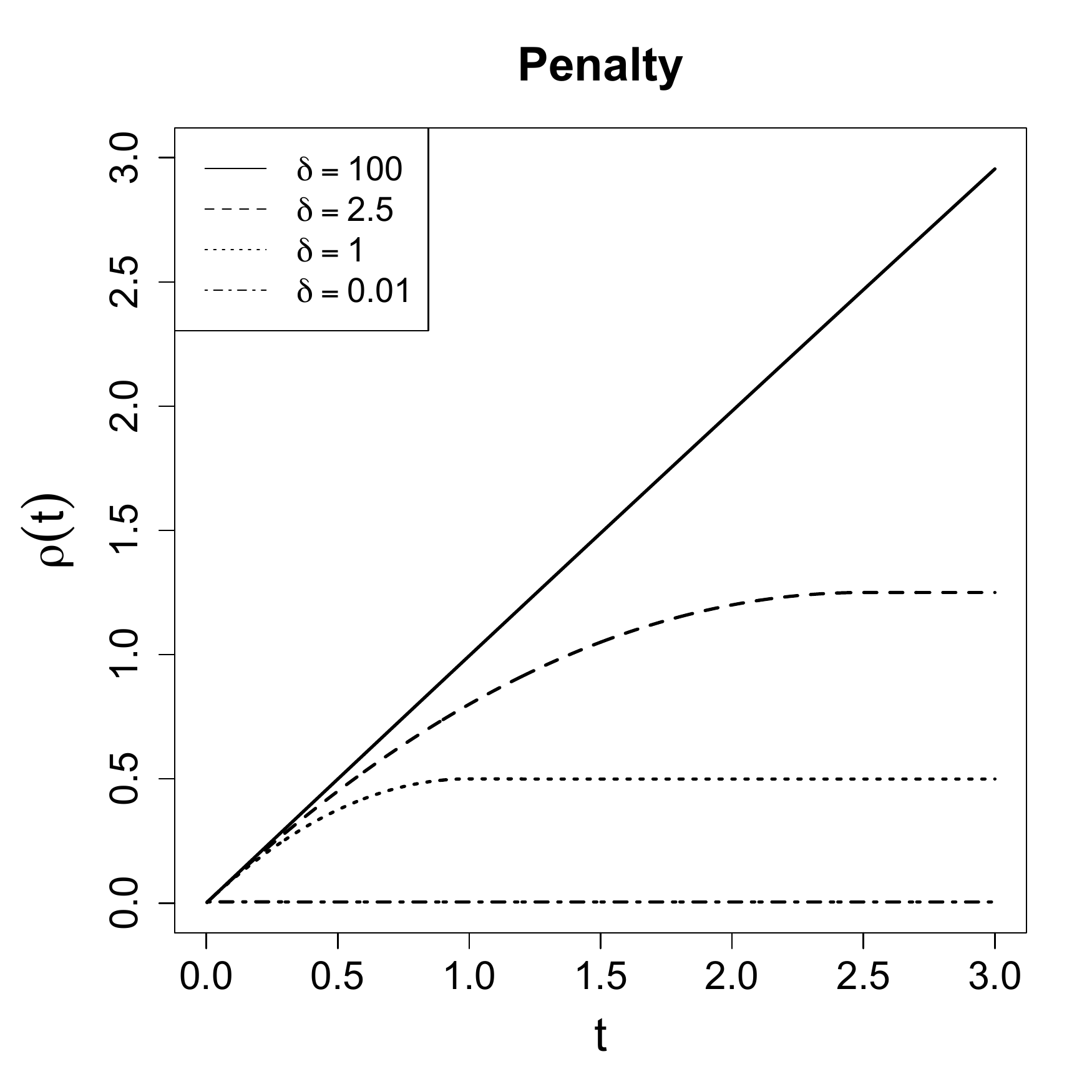}
		\caption{}
		\label{subfig11}
	\end{subfigure} %
	\begin{subfigure}[b]{0.45\textwidth}
		\centering
		\includegraphics[width=\textwidth]{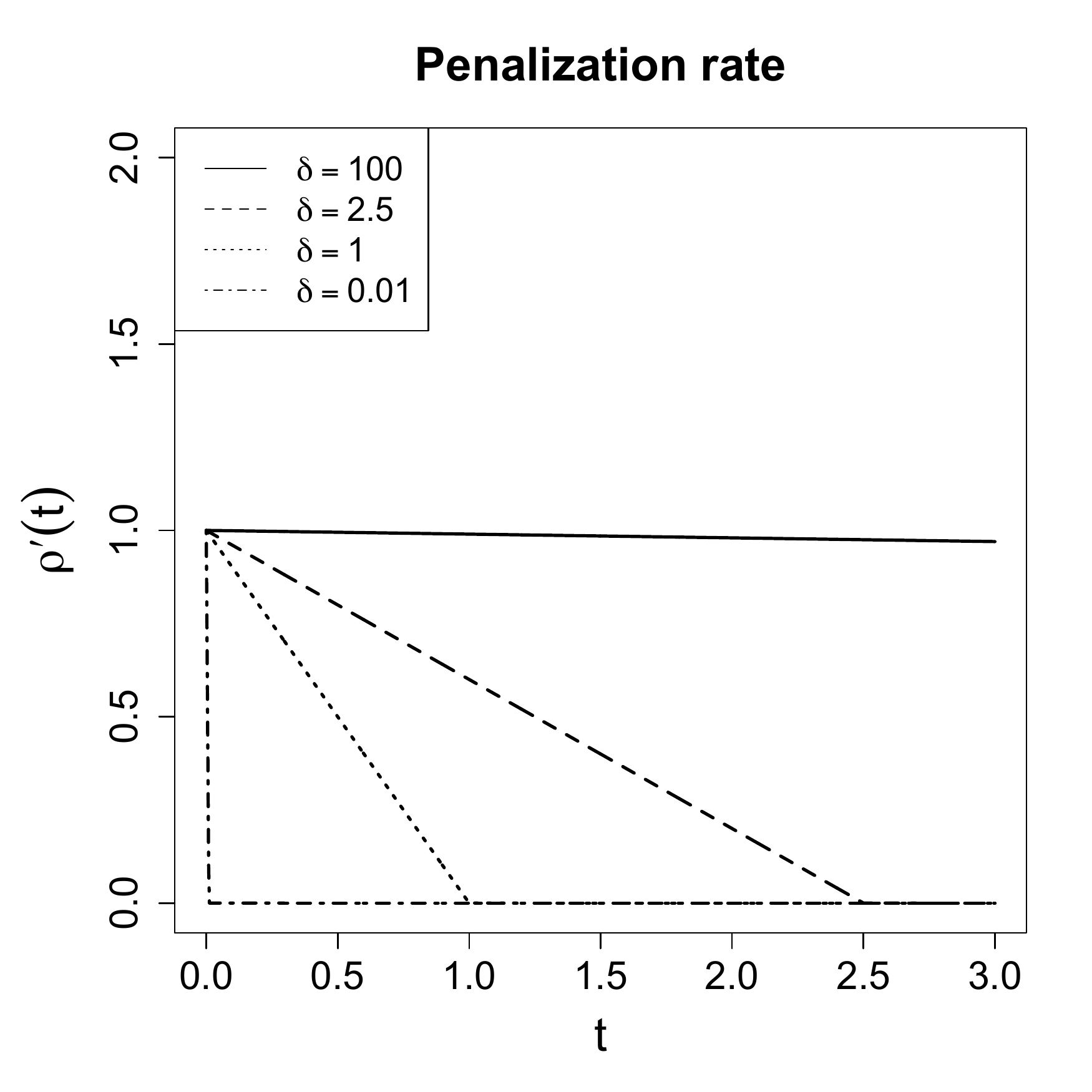}
		\caption{}
		\label{subfig12}
	\end{subfigure} %
	\caption{MCP in (a) and its derivative in (b) are plotted for different values of $\delta$ and $\lambda=1$. It approaches the $\ell_1$ penalty when $\delta$ is high (solid line).}
	\label{fig1}
\end{figure}

An appropriately chosen penalty should result in an estimator that satisfies the properties of unbiasedness, sparsity, and continuity \citep{fan2001}. To achieve these three properties and to specifically avoid excessive bias in the estimation of $\theta$, which could lead to unsatisfactory cluster assignment, we propose to use the minimax concave penalty (MCP) developed by \citet{zhang2010},
\begin{align}
	\label{eq:penalty}
	 \rho \left( t \right) & = \int_0^{t} \left( 1 - \frac{x}{\delta \lambda}  \right)_+ dx \\
	& = \left( t - \frac{t^2}{ 2 \lambda \delta }  \right) I( t < \lambda \delta ) + \left( \frac{ \lambda \delta }{ 2 } \right) I( t \geq \lambda \delta ), \quad ( t \geq 0 ), \notag
\end{align}
where $I(\cdot)$ is the indicator function.
 The MCP penalty $\rho(t)$ in \eqref{eq:penalty} defines a family of penalty functions that are concave in $t \in [ 0, \infty )$, where $\lambda > 0$ controls the amount of regularization and $\delta >0$ controls the degree of concavity. It has been noted that such non-convex penalties promote sparser models than the $\ell_1$ penalty with the same or superior prediction accuracy in regression models \citep{zhang2010, mazumder2011}. In fact, MCP includes both the $\ell_1$ penalty when $\delta \rightarrow \infty$ and the $\ell_0$ penalty when $\delta \rightarrow 0+$, forming a continuum between the two extremes. MCP is a simple differentiable penalty function with only two parameters and is designed to minimize maximum concavity. Compared to other non-convex penalties such as SCAD \citep{fan2001} or group truncated Lasso penalty \citep{pan2013}, it includes the explicit concavity parameter $\delta$ in its formulation that is easily separated from the penalization rate. Increasing the concavity through this parameter allows us to effectively control the bias of $\theta$ using a data-driven approach. The minimax concave penalty is demonstrated in Figure \ref{fig1}. 

We illustrate this regularization in the penalized loss function \eqref{eq:genloss} with a special case $n=2$. Denote the sample mean of the two observations by $\bar{y} = \frac{1}{2} (y_1 + y_2)$. The objective function in this special case is
\begin{align}
	\label{eq:twoloss}
	& \ell( \theta_1,\theta_2 ) =  \| y_1 - \theta_1 \rVert_2^2 + \| y_2 - \theta_2 \rVert_2^2 + \lambda \rho \left( \| \theta_1 - \theta_2 \|_2 \right).
\end{align}
Let $\gamma = ( \theta_2 - \theta_1 ) \in \mathbb{R}^p$. Then, for any fixed $\gamma$, $\ell( \theta_1,\theta_2 )$ is minimized at $(\theta_1, \theta_2 ) = \left( \bar{y} - \gamma/2, \bar{y} + \gamma/2 \right)$, and thus, minimizing \eqref{eq:twoloss} reduces to
\begin{align}
	 \min_{\gamma} \biggl[ \ell( \gamma ) & =   \| y_1 - \bar{y} + \gamma/2 \rVert_2^2 + \| y_2 - \bar{y} - \gamma/2 \rVert_2^2 + \lambda \rho \left( \| \gamma \|_2 \right)  \notag \\
	& = \frac{1}{2} \| \gamma - ( y_2 - y_1 ) \|_2^2 + \lambda \rho \left( \| \gamma \|_2 \right) \biggr].\label{eq:twoloss_gamma}
\end{align}

\begin{figure}[ht]
	\centering
	\begin{subfigure}[b]{0.45\textwidth}
		\centering
		\includegraphics[width=\textwidth]{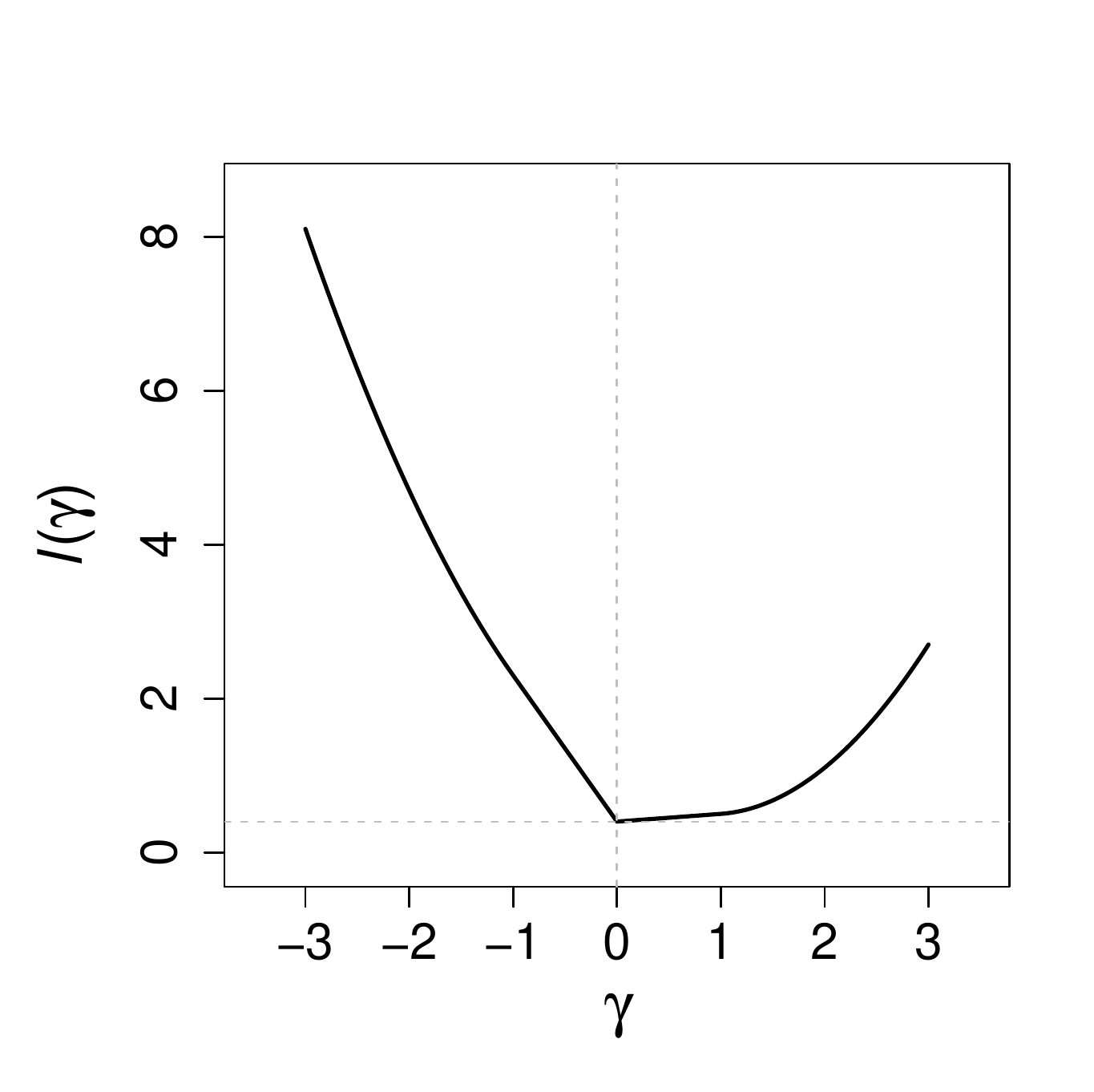}
		\caption{}
		\label{subfig21}
	\end{subfigure} %
	\begin{subfigure}[b]{0.45\textwidth}
		\centering
		\includegraphics[width=\textwidth]{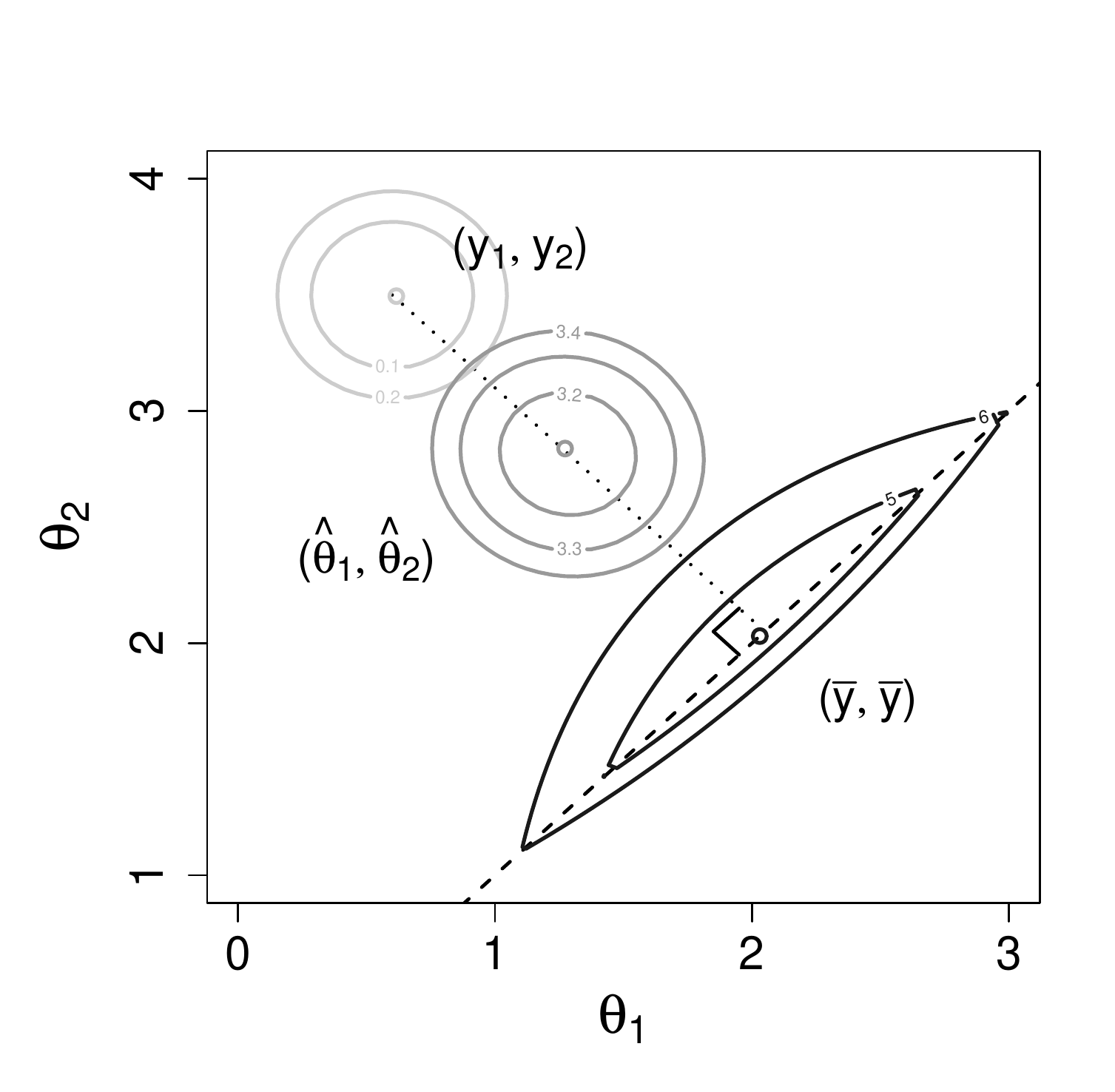}
		\caption{}
		\label{subfig22}
	\end{subfigure} %
	\caption{Demonstration of the regularization with $n=2$ and $p=1$. (a) The penalized loss function $\ell( \gamma )$ with $\lambda=1$ and $\delta=1$. (b) The contours of the unpenalized (light gray) and penalized (dark gray and black) $\ell_2$ loss as a function of $\theta_1$ and $\theta_2$. The dark gray contours centered at $(\hat{\theta}_1, \hat{\theta}_2) = (1.3, 2.8)$ demonstrate the bias in the estimates of the cluster centers.}
	\label{fig2}
\end{figure}

Figure \ref{subfig21} plots the objective function $\ell( \gamma )$ for $p=1$. It can be seen from the figure that $\ell( \gamma )$ is minimized when $\gamma = 0$, i.e. $\theta_1 = \theta_2 = \bar{y}$ for the chosen $\lambda$. Figure \ref{subfig22} depicts how the penalty imposed on the difference $| \theta_1 - \theta_2 |$ helps obtain $\hat{\theta}_1 = \hat{\theta}_2$. The unpenalized $\ell_2$ loss function has circular light gray contours centered at $(y_1, y_2)$ and the penalized $\ell_2$ loss function's contour is drawn in black with the minimum at $(\bar{y}, \bar{y})$. The penalty on $|\theta_1 - \theta_2|$ forces the minimizer to move from $(y_1, y_2)$, when the loss function has no penalty, to $\hat{\theta}_1 = \hat{\theta}_2 = \bar{y}$, which lies on the dashed line $\theta_1 = \theta_2$.

In Figure~\ref{subfig22} we also demonstrate how the choice of $\delta$ and $\lambda$ for the MCP in \eqref{eq:penalty} affects the estimates of the cluster centers. The light gray contours  in Figure~\ref{subfig22} represent the unpenalized $\ell_2$ loss (with $\delta \rightarrow 0$ or $\lambda \rightarrow 0$), which is minimized at $( y_1, y_2 )$ and gives two different clusters. The black contours in the same figure depict the objective function \eqref{eq:twoloss} when the value of $\delta$ is low enough and the value of $\lambda$ is big enough to produce an unbiased estimate of one cluster center $\hat{\theta}_1 = \hat{\theta}_2 = \bar{y}$. The dark gray contours plot the objective function for a larger value of $\delta$ and a smaller value of $\lambda$, which is minimized at $(\hat{\theta}_1, \hat{\theta}_2) = (1.3,2.8)$. In this case, two clusters are obtained, and both centers are estimated with substantial bias. Therefore, a proper and data-driven choice of $\delta$ and $\lambda$ is key to our method. This also shows that a penalty in the form $\| \theta_i - \theta_j \|_2$, which corresponds to $\delta \rightarrow \infty$, may not be appropriate for clustering.

\section{Solution path clustering}
\label{spc}

\subsection{An MM algorithm}
\label{mm}

Minimization of a non-convex objective function is usually non-trivial. Motivated by the MM algorithm \citep{lange2004} we propose to majorize the penalty term of \eqref{eq:genloss}  by a linear function \citep{wu2008}. We then minimize the majorizing surrogate function by cyclic block coordinate descent. We initialize the algorithm assuming all objects form singleton clusters and gradually merge the objects into a decreasing number of clusters for an appropriately chosen sequence of parameters $( \delta, \lambda )$, stopping when all the objects form one cluster. Correspondingly, once two objects are merged into a cluster, we do not consider splitting them in the later stages of the algorithm (Remark~\ref{splitting}). Suppose that in the current solution the objects $y_i$ are assigned to $K$ clusters with centers $\mu_1, \ldots, \mu_K$, 
where $\mu_k \in \mathbb{R}^p$. Let $C_k = \left\{ i: \theta_i = \mu_k \right\}$ represent the $k$th current cluster 
and $N_k = | C_k |$ denote the size of this cluster. 
We may then re-write the objective function \eqref{eq:genloss} as
\begin{align}
	\label{eq:kloss}
	& \ell_K( \mu ) = \sum_{k=1}^K \sum_{ i \in C_k } \| y_i - \mu_k \|_2^2 + \lambda \sum_{ k < \ell } N_k N_{\ell} \rho \left( \| \mu_k - \mu_{\ell} \|_2 \right),
\end{align}
where  $\mu = \left( \mu_1, \ldots, \mu_K \right)$.
With a proper choice of $(\lambda, \delta)$, minimizing $\ell_K( \mu )$ over $\mu$ will force some $\mu_k$'s to be very close to each other, effectively merging these clusters into a bigger cluster in the next solution.

To minimize $\ell_K( \mu )$, we employ a blockwise MM step to cycle through $\mu_k$. At each step, we fix $\mu_{[-k]} = ( \mu_1, \ldots , \mu_{k-1}, \mu_{k+1}, \ldots, \mu_{K} ) $ to its current value and majorize
\begin{align}
	\label{eq:stepkloss}
	& \ell_K( \mu_k ) \triangleq \sum_{ i \in C_k } \| y_i - \mu_k \|_2^2 + \lambda N_k \sum_{ \ell \neq k } N_{\ell} \rho \left( \| \mu_k - \mu_{\ell} \|_2 \right).
\end{align}
Let $\mu_k^{(t)}$ be the value of $\mu_k$ before the current MM step, where $t$ corresponds to the iteration number. By assumption $\mu_k^{(t)} \neq \mu_{\ell}$ for all $\ell$, and we majorize $\rho \left( \| \mu_k - \mu_{\ell} \|_2 \right)$ by
\begin{align}
	\label{eq:majorization}
	  & \rho \left( \| \mu_k - \mu_{\ell} \|_2 \right)  \\
	  \leq &  \rho \left(  \| \mu_k^{(t)} - \mu_{\ell} \|_2 \right) + \rho^{\prime} \left(  \| \mu_k^{(t)} - \mu_{\ell}  \|_2 \right) \left( \| \mu_k - \mu_{\ell} \|_2 -  \| \mu_k^{(t)} - \mu_{\ell} \|_2 \right) \notag \\
	 \leq & \rho \left(  \| \mu_k^{(t)} - \mu_{\ell} \|_2 \right) + \rho^{\prime} \left(  \| \mu_k^{(t)} - \mu_{\ell} \|_2 \right) \left(   \frac{  \| \mu_k - \mu_{\ell} \|_2^2 - \| \mu_k^{(t)} - \mu_{\ell} \|_2^2 }{  2 \| \mu_k^{(t)} - \mu_{\ell} \|_2 } \right), \notag
\end{align}
due to the concavity of the functions $\rho(x)$ and $\sqrt{x}$ for $x>0$. When the majorization \eqref{eq:majorization} is substituted into $\ell_K( \mu_k )$ \eqref{eq:stepkloss} we obtain a quadratic surrogate function in $\mu_k$ for which the minimizer is
\begin{align}
	\label{eq:mu}
	& \mu_k^{(t+1)} = \frac{ \bar{y}_k + \lambda \sum_{ \ell \neq k } w_{k,\ell}^{(t)}  \mu_{\ell} }{  1 + \lambda \sum_{ \ell \neq k } w_{k,\ell}^{(t)} },
\end{align}
where $\bar{y}_k = \frac{1}{N_k}\sum_{i \in C_k } y_i$ and $w_{k,\ell}^{(t)}$ can be regarded as the weight for $\mu_{\ell}$:
\begin{align}
	\label{eq:weight}
	& w_{k,\ell}^{(t)} =  \frac{ N_{\ell} \rho^{\prime} \left(  \| \mu_k^{(t)} - \mu_{\ell}  \|_2 \right) } { 2 \| \mu_k^{(t)} - \mu_{\ell} \|_2  }=  \frac{ N_{\ell }\left( 1 - \| \mu_k^{(t)} - \mu_{\ell} \|_2/\lambda \delta \right)_{+} }{ 2 \| \mu_k^{(t)} - \mu_{\ell} \|_2 }.
\end{align}
In effect, the derivative $\rho^{\prime}$ in \eqref{eq:weight} becomes the adaptive weight for the estimation of the cluster centers $\mu_k$, which is similar to the convex clustering penalty weights in \citet{hocking2011} and \citet{chi2013}. The weight in \eqref{eq:weight}, however, uses the distances between the cluster centers $\mu_k$ and varies with each iteration, whereas weights in \citet{hocking2011} and \citet{chi2013} are based on the distances between the data points, which do not change throughout the estimation procedure and a solution path. It can be seen from \eqref{eq:mu} and \eqref{eq:weight} that:
\begin{itemize}
	\item When $\| \mu_k^{(t)} - \mu_{\ell} \|_2 \geq \lambda \delta$ for all $\ell \neq k$, then $w_{k,\ell}^{(t)} = 0$ and, thus, in the next iteration $\mu^{(t+1)}_k = \bar{y}_k$, the sample mean.
	\item When $\| \mu_k^{(t)} - \mu_{\ell} \|_2 \ll \lambda \delta$ so that $\lambda w_{k,\ell}^{(t)} \gg 1$ for a particular $\ell$, then $\mu^{(t+1)}_k \approx \mu_{\ell}$ and the two clusters $C_k$ and $C_{\ell}$ will merge.
\end{itemize}

A single iteration of the MM algorithm cycles through all $K$ blocks as summarized in Algorithm~\ref{alg:MM}.
To account for data scaling we set $\xi= \frac{ \epsilon }{ \sqrt{p} } \sum_{m=1}^p \sigma_m$, where  $\epsilon = 10^{-4}$ and $\sigma_m$ is the standard deviation of the $m$th component of the data. 

\algsetup{indent=1.5em}
\begin{algorithm}
\caption{One iteration of the MM algorithm}
\label{alg:MM}
\begin{algorithmic}[1]
\FOR{$k=1,\ldots,K$}
\STATE{majorization: compute weights $w_{k,\ell}^{(t)}$ as in \eqref{eq:weight} for all $\ell \neq k$ }
\STATE{minimization: update $\mu_k^{(t+1)}$ as in \eqref{eq:mu} }
\IF{$\| \mu_k^{(t+1)}  -  \mu_{\ell} \|_2 < \xi$ for some $\ell$} \label{alg:MM:merge_st} 
\STATE{set $\mu_k^{(t+1)}$ and $\mu_{\ell}$ to their weighted mean}
\ENDIF \label{alg:MM:merge_end}
\ENDFOR
\end{algorithmic}	
\end{algorithm}

\begin{remark}[Relative sparsity]
\label{relsparsity}
It should be noted that the minimizers of \eqref{eq:kloss} in general do not have exactly identical pairs
of $\mu_k$'s for any finite $\lambda$, and thus only relative sparsity can be achieved so that some $\mu_k$'s become very close to each other. This is due to the penalized loss function \eqref{eq:kloss} itself, not because of the algorithm implementation. However,
such relative sparsity is sufficient for practical applications, with 
a simple thresholding step like on lines~\ref{alg:MM:merge_st}-\ref{alg:MM:merge_end} of Algorithm \ref{alg:MM}. We perform the thresholding step for every iteration of the MM algorithm to decrease the overall running time of SPC as $K$ may become smaller for the next iteration. 
\end{remark}

\begin{remark}[Cluster splitting]
\label{splitting}
We have made an assumption that a cluster is never split so that $\mu_k$'s are gradually merged into a decreasing number of clusters. This assumption is made mainly to cut down on the computation time and it has worked well in practice. However, in order to exactly solve the problem in \eqref{eq:genloss} it is necessary that objects are allowed to be split or unfused from the clusters they were assigned to. The splitting of clusters could be handled easily within the framework of our algorithm by an additional soft thresholding step. The details on cluster splitting by soft thresholding are provided in the Appendix.
\end{remark}

\begin{remark}[Convergence]
Convergence for coordinate descent type algorithms has been established for the sum of a smooth function and a non-convex penalty under certain conditions that are in fact met by MCP \citep{mazumder2011}. However, since the penalty term in \eqref{eq:stepkloss} is non-separable, this result as well as that in \citet{tseng2009} cannot be applied. Another difficulty is that the objective function \eqref{eq:stepkloss} is non-differentiable when $\mu_k=\mu_{\ell}$, which does not meet the assumption for the fixed points of the MM algorithm to coincide with the set of the stationary points of the objective function \citep{lange1995, lange2000}. Consequently, there is no theoretical guarantee that Algorithm~\ref{alg:MM} always converges to a stationary point. In our implementation, Algorithm~\ref{alg:MM} is repeated until it reaches convergence in the cluster centers, i.e. 
$$\displaystyle\max_{1 \leq k \leq K} \| \mu_k^{(t+1)} - \mu_k^{(t)} \|_2 < \xi,$$ 
or until 50 iterations are reached. In practice, we have observed that Algorithm~\ref{alg:MM} almost always converges in fewer than 50 iterations. See Section~\ref{simstudy} for empirical evidence for the convergence of our algorithm.
\end{remark}

\subsection{Solution path construction}
\label{solutionpath}

The penalty function in \eqref{eq:penalty} has two parameters $\lambda$ and $\delta$, the former controlling the amount of regularization and the later determining the degree of concavity of the function. We would like to create some simple data-driven rules for selecting several combinations of the penalty parameters to produce a solution path for any clustering problem. We start the algorithm assuming no sparsity and that each individual observation $y_i$ forms its own singleton cluster. We then gradually enforce greater sparsity by using an increasing sequence of $\lambda$ while reducing the bias, if necessary, through a decreasing sequence of $\delta$. Since the penalty function \eqref{eq:penalty} depends on the distances between $\mu_k$'s, the sequences of the tuning parameters can be guided by these distances, and the solution from the current combination of $(\delta, \lambda)$ can be used as a warm start for the next combination.

We define a decreasing sequence $\Delta =\{ \delta_1, \ldots, \delta_H \} $ and for each $\delta_h$, $h=1,\ldots, H$, define an increasing sequence $\Lambda( \delta_h ) = \{ \lambda_1( \delta_h ), \ldots, \lambda_G ( \delta_h ) \}$. The sequence $\Delta$ is simply determined by 
\begin{align}
	\label{eq:delta}
	& \delta_{h} = \delta_{h-1} \alpha, 
\end{align}
for $h=2, \ldots, H$, where $\alpha \in ( 0, 1 )$ is a constant. We discuss the choice of $\delta_1$ a little later in this section when we talk about $\lambda_1 (\delta_1)$. Each time the value of $\delta$ is decreased a new $\Lambda ( \delta )$ is computed. The initial solution is then obtained with the lowest concavity. As mentioned previously, high values of $\delta$ decrease concavity of the penalty function and make it behave more like the $\ell_1$ penalty which could introduce considerable bias into the estimate of $\mu_k$. Therefore, in order to determine whether the value of $\delta$ needs to be decreased we define the bias-variance ratio (BVR) for each cluster $C_k$ as
\begin{align}
	\label{eq:bvr}
	& \text{BVR}_k
	= \left\{
	\begin{array}{lll}
		\frac{\| \mu_k - \bar{y}_k \|_2^2}{ \sum_{i \in C_k} \| y_i - \bar{y}_k \|_2^2 / ( N_k - 1 ) }, & N_k > 1 \\
		\\
		\frac{\| \mu_k - y_i \|_2^2}{ \left(r_k/ 2 \right)^2 }, & N_k = 1, C_k = \{ i \},
	\end{array} \right.
\end{align}
where $r_k=\min_{\ell \ne k}\| y_i - \mu_{\ell}\|_2$ is the distance between $y_i$ and the nearest cluster center.  If $\text{BVR}_k > 1$ for any $k$, then we decrease $\delta$ as in \eqref{eq:delta}. The idea behind the bias-variance ratio is that if a certain estimated cluster center $\mu_k$ moves beyond the range of the observations $y_i$ in that cluster $C_k$, then the concavity is increased in order to reduce the bias that could lead to a bad solution.

We now address the choice of $\Lambda(\delta_h)$ determined by the lower and upper bounds of the sequence, $\lambda_1 ( \delta_h )$ and $\lambda_G (\delta_h)$. The values of $\lambda$ in between are evenly spaced in log-scale. We state two lemmas for a simple case with $n=2$ and then use these lemmas to motivate our choice of the lower and upper bounds for $\Lambda(\delta_h)$. The proofs for the lemmas are provided in the Appendix.
\begin{lemma}\label{lemma1}
Assume $n=2$ and that there are only two points $(y_1, y_2)$ with distance $d = \| y_1 - y_2 \|_2$. 
Let $\theta_i^{(t)}$ be the value of $\theta_i$ generated by the MM algorithm with $\theta_i^{(0)}=y_i$, $i=1,2$.
Fix $\lambda \delta = \eta>0$. For a given $\phi \in (0, 1)$, if $\eta >d$ and 
\begin{align}
\label{eq:lemma1}
& \lambda = \frac{2 \phi \eta d }{ \left( 1 - \phi \right) \left( \eta - d \right) },
\end{align}
then $\| \theta_1^{(1)} - \theta_2^{(0)} \|_2 = ( 1 - \phi ) \| \theta_1^{(0)} - \theta_2^{(0)} \|_2 $. 
If $\eta \leq d$, then $\theta_i^{(t)}=y_i$ for all $t\geq 1$ and $i=1,2$.
\end{lemma}
\begin{lemma}
\label{lemma2}
Assume $n=2$ and that there are only two points $(y_1, y_2)$ with distance $d = \| y_1 - y_2 \|_2$. For a given $\delta > 0$, if 
\begin{align}
& \lambda  \geq \left( 1 + \frac{ 1 }{ \delta } \right) d,
\label{eq:lemma2}
\end{align}
then the global minimizer of \eqref{eq:twoloss} is given by $( \hat{\theta}_1, \hat{\theta}_2 ) = ( \bar{y}, \bar{y} )$.
\end{lemma}

We use Lemma~\ref{lemma1} to determine $\delta_1$ and the initial lower bound $\lambda_1(\delta_1)$. 
From this lemma one sees that $\eta$ serves as a threshold: if $d\geq \eta$, then $\theta^{(t)}_i$ will not change
and the points will not merge. 
Let $\eta_1 = \lambda_1(\delta_1) \delta_1$ and $Q_{\beta}$ be the $\beta$-quantile of the nearest neighbor distances among $y_i$'s. We choose $\eta_1 = Q_{\omega}$, where $\omega \in (0,1)$ can be regarded as the approximate proportion of data points that may merge in the initial solution. On the other hand,
it follows from Lemma~\ref{lemma1} that $\phi \in (0,1)$ can be considered the minimization step size. 
By default, we set $\phi=0.5$. Then, to use \eqref{eq:lemma1} to determine $\lambda_1(\delta_1)$ we need to specify $d$.
We may choose $d$ as the distance between a pair of points such that $d< \eta_1$. This can
be achieved by simply setting $d=Q_{\tau}$, where $\tau \in (0, \omega )$. 
Plugging these choices of parameters into \eqref{eq:lemma1}, we obtain the initial lower bound for $\lambda$ as
\begin{align}
	\label{eq:lambda1_1}
	& \lambda_1 ( \delta_1 ) = \frac{2 \phi Q_{\omega} Q_{\tau} }{ \left( 1 - \phi \right) \left( Q_{\omega} - Q_{\tau} \right) }.
\end{align}
The first value of the sequence $\Delta$ then follows directly from
\begin{align}
	\label{eq:delta1}
	& \delta_1 = Q_{\omega}/\lambda_1( \delta_1 ).
\end{align}

Denote the maximum penalty by $z = \lambda \max_t \rho( t ) = \frac{1}{2}\lambda^2 \delta $. In order to achieve gradual merging of objects into fewer clusters, both the threshold $\eta = \lambda \delta$ and $z$ should be non-decreasing. Suppose that $\delta_{h-1}$ is decreased to $\delta_h$ by the BVR criterion and $z_{h-1} = \frac{1}{2} \left[ \lambda_{\tilde{G}} (\delta_{h-1}) \right]^2 \delta_{h-1}$, where $\lambda_{\tilde{G}} (\delta_{h-1})$ is the  value of $\lambda$ before decrease. The lower bound for $\Lambda (\delta_h)$ is then
\begin{align}
	\label{eq:lambda1_h}
	& \lambda_1 ( \delta_{h} ) = \left( \frac{ 2 z_{h-1} }{ \delta_{h}  } \right)^{1/2} = \left( \frac{ \left[ \lambda_{\tilde{G}} (\delta_{h-1}) \right]^2 \delta_{h-1} }{ \delta_{h}  } \right)^{1/2} = \alpha^{-1/2} \lambda_{\tilde{G}} (\delta_{h-1}).
\end{align}

Next, we directly apply Lemma~\ref{lemma2} to define the upper bound $\lambda_G ( \delta_h )$, $h = 1, \ldots, H$. Given a collection of objects, we can conservatively choose $d$ in \eqref{eq:lemma2} to be the maximum distance among all pairs of objects to obtain
\begin{align}
	\label{eq:lambdaG}
	& \lambda_G ( \delta_h )  = \left( 1 + \frac{ 1 }{ \delta_h } \right) \max_{i,j} \| y_i - y_j \|_2.
\end{align}
For a collection of $n>2$ objects, the upper bound in \eqref{eq:lambdaG} becomes only an approximation for the value of $\lambda$ such that all objects merge into a single cluster. This value of $\lambda$ does not necessarily guarantee that the objects will merge but, in practice, we have not encountered a situation when this approximation did not work. In a case when the value of the upper bound of $\lambda$ is not sufficiently large, one can simply decrease the value of $\delta$, re-calculate the new sequence of $\lambda$ and run the algorithm until all data points form a single cluster.   

In summary, the construction of the solution path involves the specification of four parameters, $\alpha$ in \eqref{eq:delta}, and $\omega$, $\phi$ and $\tau$ in \eqref{eq:lambda1_1}, all ranging between 0 and 1. In effect, the parameters $\tau$ and $\phi$ control the step size of the MM iteration, and consequently the length of the solution path. Based on our experience and as demonstrated by the sensitivity analysis in Section~\ref{sensitivity}, the solution path is not much affected by the choice of $\phi$ and $\tau$, except when very small values are used ($\phi, \tau \leq 0.01$), which can slow down the progression of the solution path. We recommend setting $\tau$ to be slightly smaller than $\omega$ in order to avoid unnecessary detail in the solution path. In general, we recommend setting $\alpha=0.9$, $\phi=0.5$ and $\tau = 0.9 \omega$, and we use these default values throughout the paper. 

The only tuning parameter that needs to be specified by the user is $\omega$ for the calculation of $\lambda_1(\delta_1)$ in \eqref{eq:lambda1_1} and $\delta_1$ in \eqref{eq:delta1}. Since it stands for the approximate proportion of the nearest neighbors that may merge initially, the nature of the dataset might help determine its value. For instance, if the dataset is very noisy and clusters are not tight, $\omega$ could be set to a low value, and if the dataset is well separated into clusters and there is little noise, a high value of $\omega$ could be used.  If chosen too high in cases where the data is very noisy, $\omega$ could force the algorithm to skip the correct solution by initially merging too many noisy observations.  In practice, we also found that $\omega$ should be small in high-dimensional settings. In this paper we use $\omega = 0.1$ for the high-dimensional examples when $n<p$ and $\omega = 0.5$ when $n>p$.

\subsection{The full SPC algorithm}
\label{fullspc}

We now combine the MM algorithm with the construction of the solution path to describe the full SPC algorithm. It is initialized assuming that each observation forms its own singleton cluster and is run until all observations merge into one cluster using a sequence of $( \delta, \lambda )$. The solution obtained from a particular $( \delta, \lambda )$ is used as a warm start for the next solution. We do not require the input of the number of clusters $K$, and the algorithm typically yields a short path of 2 - 15 solutions. 

For a particular $( \delta_h, \lambda_g (\delta_h) )$, let $K( h, g )$ be the estimated number of clusters and $\hat{\mu}_k(h, g)$ for $k = 1, \ldots, K(h,g)$ be the estimated cluster centers. Using these notations, the full SPC algorithm is provided in Algorithm~\ref{alg:SPC}.

\algsetup{indent=1.5em}
\begin{algorithm}
\caption{Solution path clustering}
\label{alg:SPC}
\begin{algorithmic}
\STATE \textbf{Inputs}
\STATE{required input: $Y = (y_{im})_{n \times p}$, $\omega \in (0,1)$ }
\STATE{default input: $\tau = 0.9 \omega$, $\phi = 0.5$, $\alpha = 0.9$, $G = \min \left( 20, p \right)$ }
\STATE{initialization: $h=1$, $K=n$, $\mu_k = y_k$, $k = 1, \ldots, n$}
\end{algorithmic}
\bigskip

\begin{algorithmic}[1]
\REPEAT
\STATE{compute $\delta_h$, $\lambda_1(\delta_h)$, $\lambda_G(\delta_h)$ and construct $\Lambda(\delta_h)$ in logarithmic scale of size $G$} \label{alg:SPC:newdelta}
\FOR{ $g=1, \ldots, G$ }
\STATE{run the MM algorithm (Algorithm~\ref{alg:MM}) until $\displaystyle\max_k \| \mu_k^{(t+1)} - \mu_k^{(t)} \|_2 < \xi$ or $t > 50$ to obtain $K(g,h)$ and $\{ \hat{\mu}_k(h,g), k=1, \ldots, K(h,g)\}$} 
\FOR{$k=1, \ldots, K( h, g )$}
\STATE{compute $\text{BVR}_k$}
\IF{$\text{BVR}_k > 1$}
\STATE{ $h \leftarrow h+1$ and go to line~\ref{alg:SPC:newdelta}}
\ENDIF
\ENDFOR
\ENDFOR
\STATE{$h \leftarrow h+1$}
\UNTIL{$K( h, g )=1$}
\end{algorithmic}	
\end{algorithm}

\section{Simulation study}
\label{simstudy}

\subsection{Competing methods and cluster quality assessment}
\label{competing}

We illustrate the performance of SPC on simulated data and compare it to k-means++ algorithm \citep{arthur2007}, the convex clustering method of \citet{chi2013} and several popular clustering algorithms that were developed to account for the noise in data. For the latter purpose we selected model-based clustering (mclust) \citep{fraley2002}, tight clustering \citep{tseng2005}, and penalized weighted k-means (PWK-means) \citep{tseng2007}.

The majority of authors use the adjusted rand index (ARI) \citep{hubert1985} to compare the clustering results across different methods when the true cluster assignment is known. ARI calculates the similarity of a clustering result to the underlying true clustering assignment. It takes a maximum value of 1 if the clustering solution is identical to the true structure and takes a value close to 0 if the clustering result is obtained from random partitioning. The exact definition of the ARI can be found in the Appendix. For the simulation study we also use ARI in order to be consistent with the prevailing method of cluster quality assessment. 

Since all the competing methods, except k-means++, will detect noise, i.e. the data points that do not belong to any well defined clusters, we calculate two ARI scores $( \text{ARI}_c, \text{ARI}_n )$ for each of them. Suppose the true partition and an estimated partition are $C = \left\{ C_1, \ldots, C_R, C_{R+1}\right\}$ and $\hat{C} = \{ \hat{C}_1, \ldots, \hat{C}_K, \hat{C}_{K+1} \}$, where $C_r$ and $\hat{C}_k$ contain the indices of the data points assigned to true and estimated clusters for $r=1, \ldots, R$ and $k=1, \ldots, K$, respectively, and $C_{R+1}$ and $\hat{C}_{K+1}$ are the indices assigned to noise. Denote the respective cluster labels as $v = \left\{ v_1, \ldots, v_R, v_{R+1} \right\}$ and $\hat{u} = \left\{ \hat{u}_1, \ldots, \hat{u}_K, \hat{u}_{K+1} \right\}$, where $v_{R+1}$ and $\hat{u}_{K+1}$ indicate the labels for the true and estimated noise, respectively. See Table~\ref{table:aricounts} for the full contingency table of the counts $n_{kr} = |\hat{C}_k \cap C_r |$. Our ARI scores are calculated based on parts of the counts in this table. 

$\text{ARI}_c$ accounts for data points that are identified as belonging to estimated clusters $\hat{C}_k$, $k=1, \ldots, K$, and is calculated as in \eqref{eq:ari} in the Appendix using only the first $K$ rows of Table~\ref{table:aricounts}, $\{ n_{kr}: 1 \leq k \leq K,$ $1 \leq r \leq R+1 \}$, with the corresponding row and column sums. In effect, $\text{ARI}_c$ provides quality assessment of the identified clusters $\hat{C}_k$, i.e., misclassification of clustered data and the amount of noise in the estimated clusters.

\begin{table}[t]
\begin{center}
\caption{Contingency table and notation for the calculation of $(\text{ARI}_c,\text{ARI}_n)$}
\begin{tabular*}{0.7\textwidth}{@{\extracolsep{\fill}}c|cccc|c}
Cluster & $v_1$ & \ldots & $v_R$ & $v_{R+1}$ & Sum \\
\hline
$\hat{u}_1$ & $n_{11}$  & \ldots & $n_{1R}$ &  $n_{1(R+1)}$ & $n_{1\bullet}$ \\
\vdots & \vdots  & \vdots & \vdots & \vdots  & \vdots \\
$\hat{u}_K$ & $n_{K1}$ & \ldots & $n_{KR}$ & $n_{K(R+1)}$ &  $n_{K\bullet}$ \\
$\hat{u}_{K+1}$ & $n_{(K+1)1}$ & \ldots & $n_{(K+1)R}$ & $n_{(K+1)(R+1)}$ & $n_{(K+1)\bullet}$ \\
\hline
Sum & $n_{\bullet1}$ & \ldots & $n_{\bullet R}$ & $n_{\bullet (R+1)}$ &  $n$ \\
\end{tabular*}
\label{table:aricounts}
\end{center}
\end{table}

 $\text{ARI}_n$ indicates how sensitive a method is in identifying noise and whether any clustered data point is misclassified as noise. It is based on all the data points except the noise in the estimated clusters which is accounted for in $\text{ARI}_c$. We collapse Table~\ref{table:aricounts} to a $2 \times 2$ table with counts $n_{cc}^* = \sum_{k=1}^K \sum_{r=1}^{R} n_{kr}$, $n_{nn}^* = n_{(K+1)(R+1)}$, $n_{nc}^* = \sum_{r=1}^{R} n_{(K+1)r}$, and $n_{cn}^* = \sum_{k=1}^{K} n_{k(R+1)}$. We set $n_{cn}^* = 0$ since we account for these data points in $\text{ARI}_c$. The total number of data points to be considered for $\text{ARI}_n$ is thus $n^* = n^*_{cc} + n^*_{nn} + n^*_{nc}$. Again, we use \eqref{eq:ari} in the Appendix to calculate $\text{ARI}_n$ by plugging in $n^*_{cc}$, $n^*_{nn}$, $n^*_{nc}$, and $n_{cn}^* = 0$. 
 In general, if $n^*_{nn}$ is large and $n^*_{nc}$ is small, $\text{ARI}_n$ will be close to 1.

PWK-means, tight clustering, and k-means++, which is a classical k-means method combined with a randomized seeding procedure to select the starting centers,  require, as an input, an estimated number of clusters, and mclust requires an input of a range of the number of clusters. In most cases, we provided the comparison algorithms with ideal input parameters, which will likely result in optimal performance for these methods. In addition to the number of clusters, the competing methods, except k-means++, have other tuning parameters that we mention below. 

PWK-means also requires the input of a penalty parameter $\lambda$ since this method imposes penalty on the number of noisy data points. We used the suggested prediction-based resampling method \citep{tibshirani2005_1} to find $\lambda$, calculated the prediction strength criterion for an increasing sequence of $\lambda$, and selected the value of $\lambda$ corresponding to the highest prediction strength computed.  

Tight clustering has several tuning parameters, but most of them are recommended to stay at their default values, which we follow for the simulated data. Along with the user-specified target number of clusters $k_{\text{target}}$, tight clustering relies on a starting number of clusters $k_0 > k_{\text{target}}$. The tight clustering algorithm is then applied to a decreasing sequence, decremented by 1, starting with $k_0$ and ending with $k_{\text{target}}$. The authors recommend $k_0 \geq k_{\text{target}} + 5$, however, a too large $k_0$ results in smaller clusters and many of the clustered data identified as noise. Conversely, a small $k_0$ can result in a smaller size of the estimated noise and more noise assigned to clusters. We set $k_0 = k_{\text{target}} + 5$, the smallest value according to the suggested choice. 

One additional tuning parameter of mclust is the reciprocal of the hypervolume $V$ of the data region, and the authors note that the method is sensitive to this value. The default method of calculating the hypervolume is $V = \prod_{j=1}^p ( \max_i \{y_{ij}\} - \min_i \{y_{ij}\} )$, which is used for the simulated data. Another input into mclust is the estimated categorization of each data point as clustered data or noise. Once the categorization is provided, mclust applies hierarchical clustering to the identified clustered data to get a good initialization for the EM algorithm that generates a final clustering result. We have used the recommended $K$th-nearest neighbor cleaning method \citep{byers1998} to obtain the initial categorization into clusters and noise. It is suggested that the value of $K$ should be the size of the smallest cluster to be detected, but if $K$ is selected too high, the noise identification might not perform well. We chose $K=5$ for the nearest neighbor cleaning such that it is smaller than the average cluster size.

Finally, convex clustering requires the specification of three tuning parameters, which are the Gaussian kernel weight $\phi$, the number of nearest neighbors $k$, and an increasing sequence of the regularization parameter $\gamma$ for obtaining a solution path. We chose the parameter $\phi$ such that the penalty weights $w_{ij} = \iota^k_{\{i,j\}} \exp \left( - \phi \| y_i - y_j \|_2^2 \right)$, where $\iota^k_{\{i,j\}}$ is $1$ if $y_j$ is among $y_i$'s $k$-nearest neighbors and $0$ otherwise, are on average around $0.25$. This rule forced the parameter $\phi$ to be very close to $0$ for our simulated data but any larger values for $\phi$ resulted in unsatisfactory clustering. We also used $k=5$ and an increasing sequence of $\gamma \in [0,50]$ of size 20 at even intervals, following an example in the documentation for the R package \verb+cvxclust+. 

To compare the performance, we created four different clustering scenarios: 1) well-separated clusters, 2) overlapping clusters, 3) well-separated clusters with added noise, 4) overlapping clusters with added noise. All clusters in the examples in Sections~\ref{np} and \ref{pn} are spherical, with equal variance. The cluster centers and the noise points were generated from a uniform distribution on $[-5,5]^{p}$. All noise was generated outside of the radius of the clusters, where the radius is the largest distance from the cluster center to the data points in that cluster. Overlapping clusters were generated such that 15-20\% of the data points in a pair of clusters are located within the radiuses of both clusters. For each scenario, we simulated 20 datasets with $n > p$ and $n < p$. To demonstrate the ability of SPC to identify noise and for the purposes of comparison with other methods, we simply regard all estimated clusters of size $N_k \leq 3$ as noise. The same cut off is also used for defining noise from k-means++ and convex clustering results.

\subsection{Results for $n>p$}
\label{np}

The simulated datasets for both well-separated and overlapping scenarios when $n>p$ are of size $n=400$, dimension $p=20$, and with $K=10$ clusters. For each of the two scenarios with noise, $200$ uniformly distributed noise points were added to the clustered data. We chose $\omega = 0.5$ for all the scenarios assuming that about half of the nearest neighbors should merge. As the output from each dataset, we obtained a solution path of 7-12 solutions, each containing the number of clusters $K$ with size $N_k > 3$, estimated cluster centers $\hat{\mu}_k$, and cluster assignments $C_k$. 

\begin{figure}
	\centering
	\includegraphics[width=\textwidth]{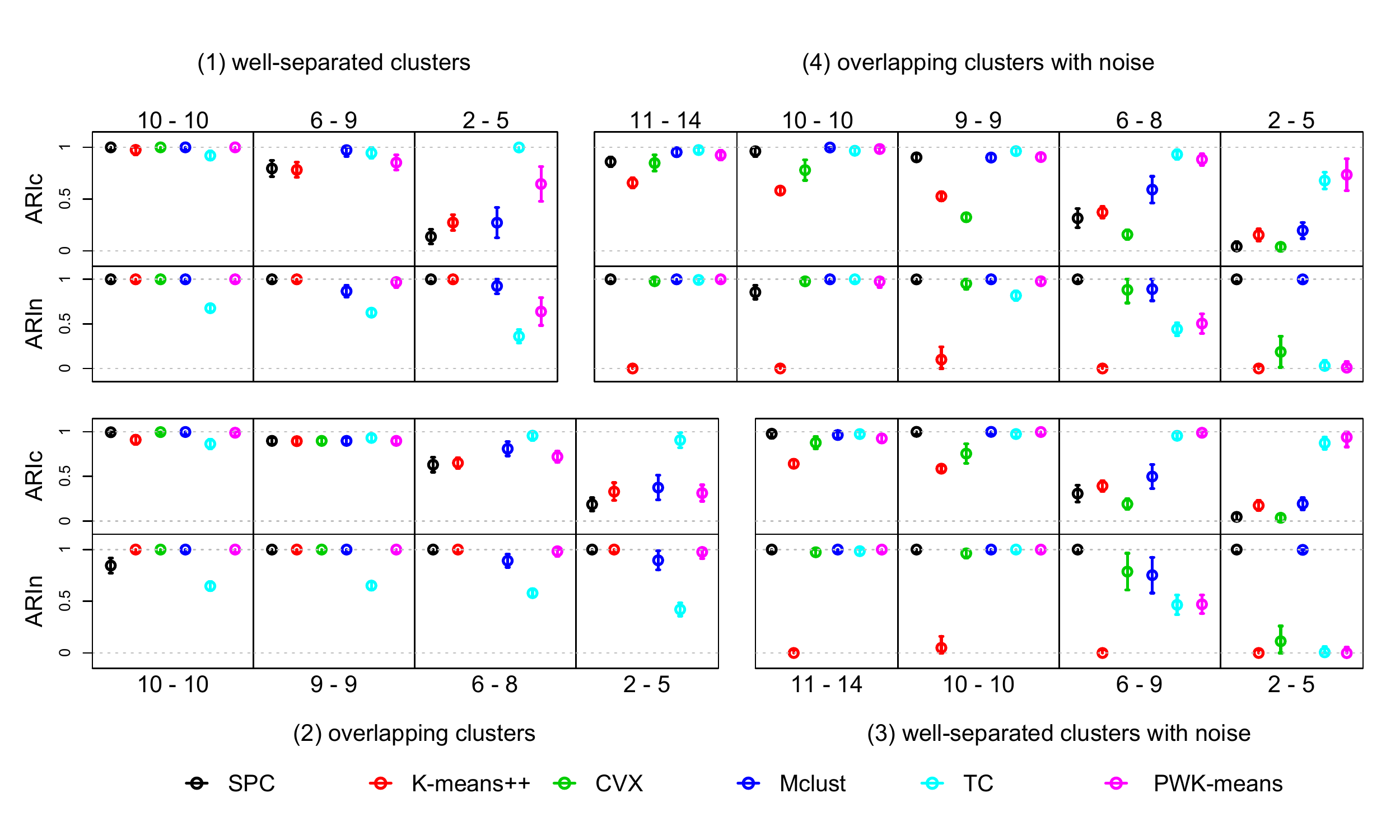}
	\caption{$\text{ARI}_c$ and $\text{ARI}_n$ for comparison methods for all four scenarios when $n>p$. Each point indicates the average $\text{ARI}_c$ (top halves of the plots) or $\text{ARI}_n$ (bottom halves of the plots) score across 20 datasets and the error bar shows the 95\% confidence interval. The range of numbers above or below each block refers to the number of clusters of size $N_k > 3$ found by SPC. For (1) and (2) $\text{ARI}_n$ reports $\text{S}_n=1-n_{nc}^*/n$. }
	\label{ARI_bign}
\end{figure}

The $\text{ARI}_c$ and $\text{ARI}_n$ scores for the comparison with other methods are presented in Figure~\ref{ARI_bign}. In addition to the true number of clusters $K=10$, we supplied the competing methods with the number of clusters (of size $N_k > 3$) along the SPC solution path to demonstrate their performance when the number of clusters is mis-specified. We report the ARI scores averaged over 20 datasets for $K=10$ and for different ranges of $K$, e.g. $6-9$, since each of the SPC solution paths for the 20 datasets might contain a different number of clusters. It must be noted that $\text{ARI}_n = 0$ when no noise is detected, which is misleading for the scenarios without noisy data. Thus, we use a special score, $\text{S}_n = 1-n_{nc}^*/n$, for these scenarios, such that if no noise is identified by a method, then $\text{S}_n = 1$. SPC can clearly outperform k-means++, especially in the scenarios with noise, due to the fact that k-means++ is not designed to separate noisy observations. SPC performs similarly or slightly better in most scenarios compared with tight clustering and PWK-means. For $K=10$, mclust outperforms all the other methods in all scenarios and separates the overlapping clusters and noise well. The pre-classification of the noise and hierarchical clustering of the remaining data provide mclust with an excellent initialization. The spherical nature of the clusters and uniformly generated noise also perfectly match the model assumptions of mclust. For $K=10$ or a slightly smaller value, the performance of SPC is very comparable to that of mclust for noisy data scenarios, although our method makes much weaker data generation assumptions and does not use any specific initialization.

All the methods recover well-separated clusters with accuracy whether noise is present or not, however, tight clustering tends to considerably underperform when no noise is present due to the fact that it is targeted specifically for noisy data. When clusters are not well-separated, SPC tends to merge overlapping clusters but identifies noise very well (high $\text{ARI}_n$). With the progression toward greater sparsity (smaller $K$), SPC adds more and more noise into the clusters or creates bigger clusters from noise, which results in a low $\text{ARI}_c$ score. On the other hand, the remaining noise is identified accurately, which is reflected in the high $\text{ARI}_n$ scores. In contrast, when the number of clusters is mis-specified and $K < 10$, tight clustering and PWK-means have the tendency to leave out clustered data as noise, reflected by low $\text{ARI}_n$ scores, while mclust merges the closest clusters together. Convex clustering performs well with overlapping clusters, however, it tends to produce less satisfactory results when any noise is present, compared to SPC. In the scenarios with noise, aside from k-means++, convex clustering adds the most noise into the clusters.

\begin{figure}
	\centering
	\begin{subfigure}[b]{0.32\textwidth}
		\centering
		\includegraphics[width=\textwidth]{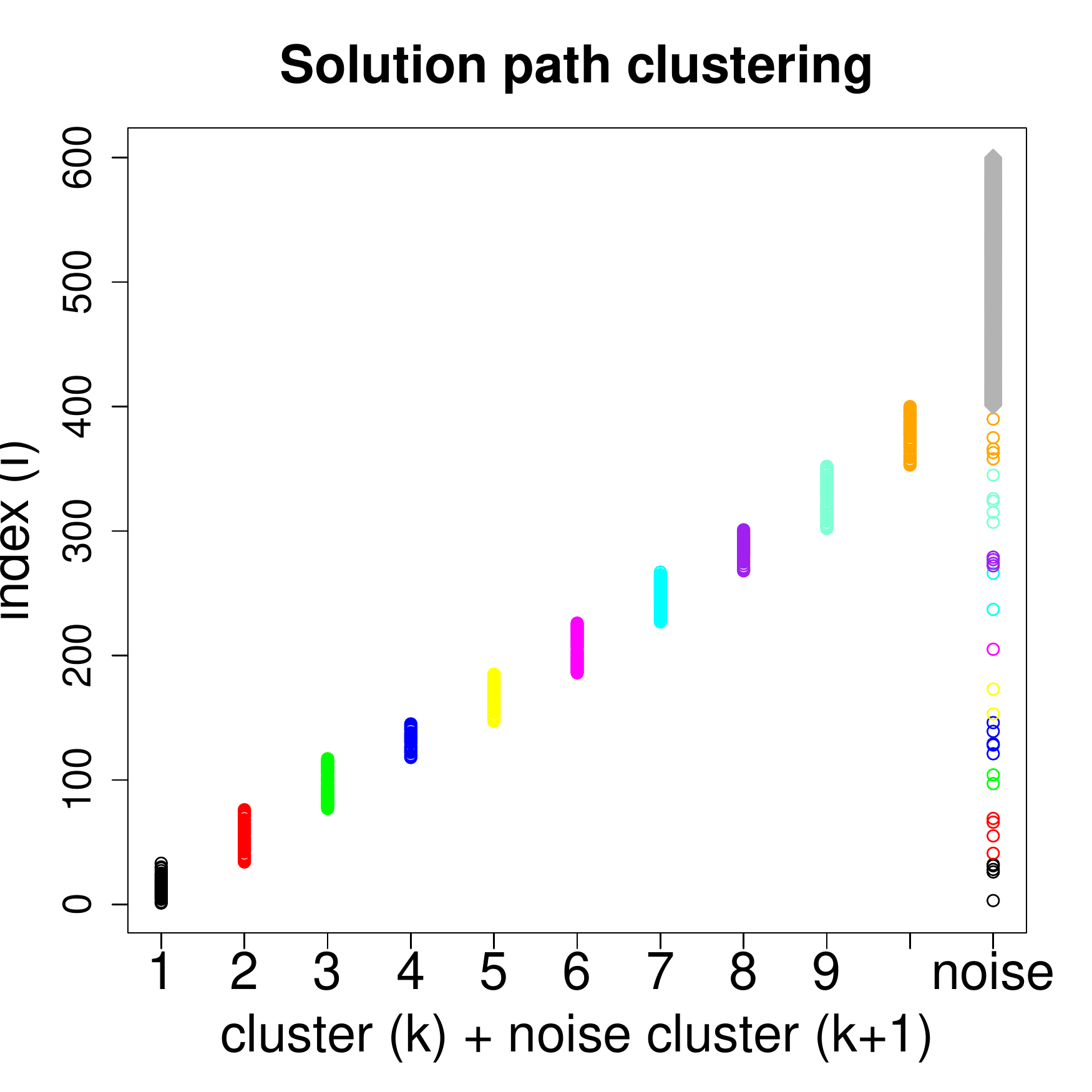}
		\caption{}
		\label{subfig41}
	\end{subfigure} %
	\begin{subfigure}[b]{0.32\textwidth}
		\centering
		\includegraphics[width=\textwidth]{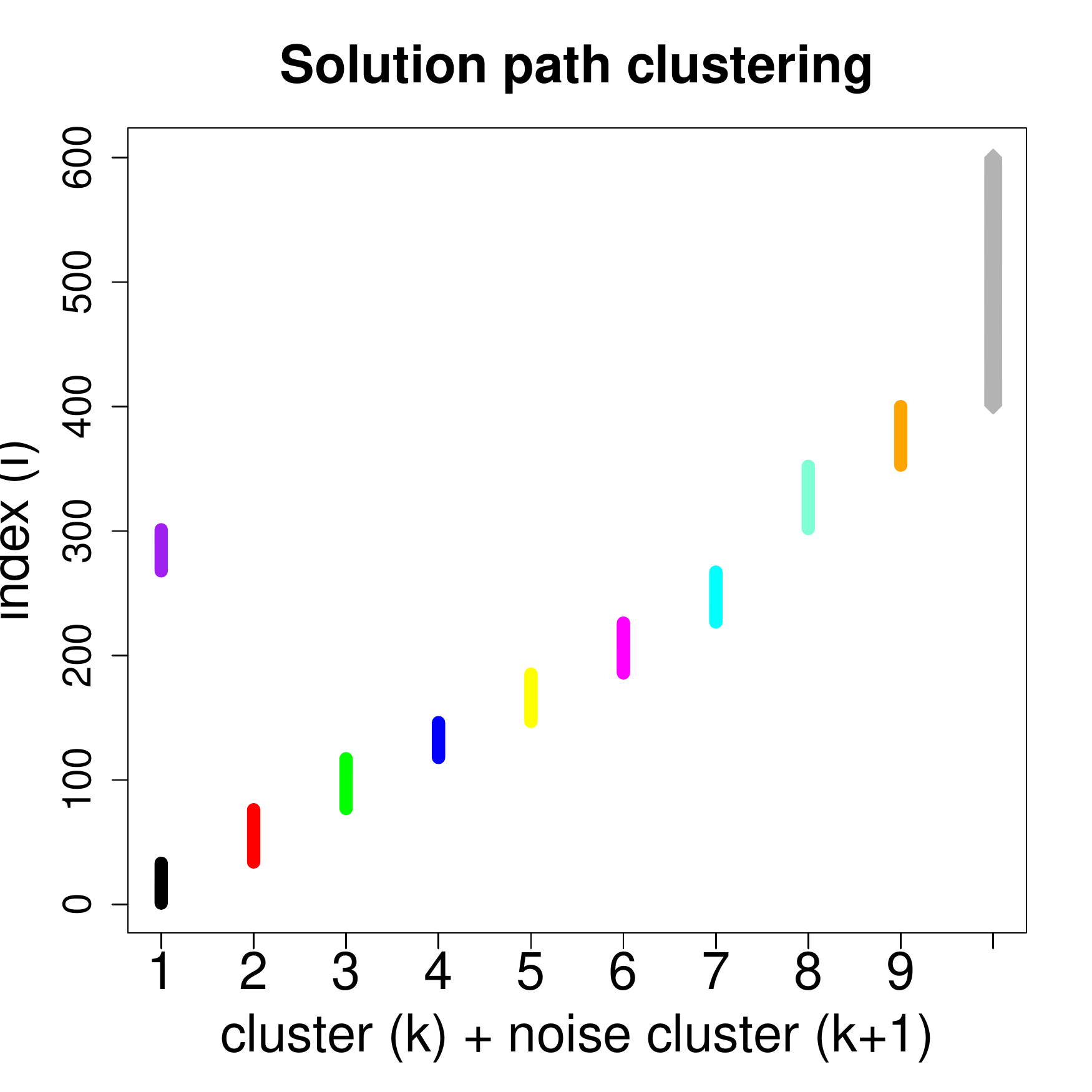}
		\caption{}
		\label{subfig42}
	\end{subfigure} %
	\begin{subfigure}[b]{0.32\textwidth}
		\centering
		\includegraphics[width=\textwidth]{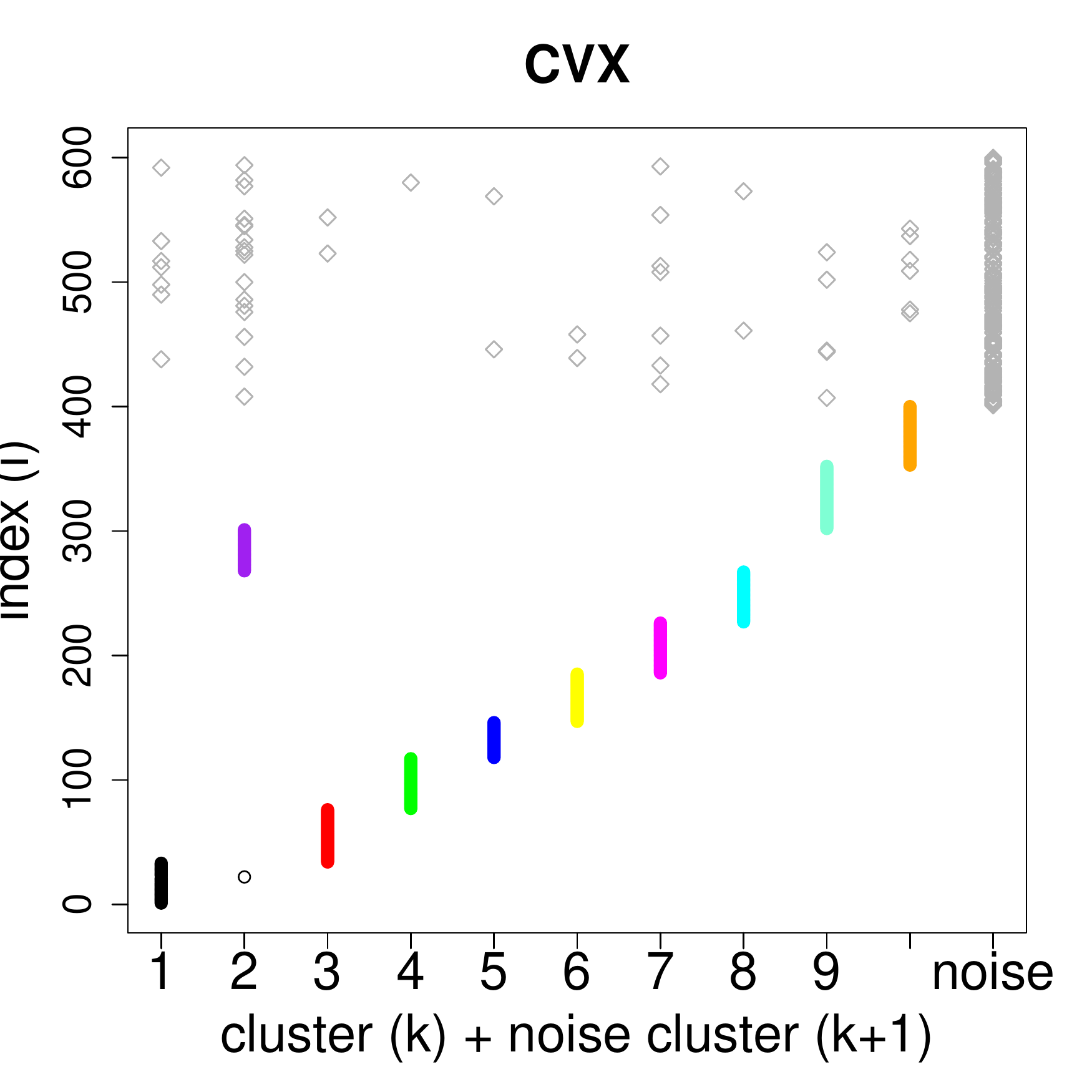}
		\caption{}
		\label{subfig43}
	\end{subfigure}
	\begin{subfigure}[b]{0.32\textwidth}
		\centering
		\includegraphics[width=\textwidth]{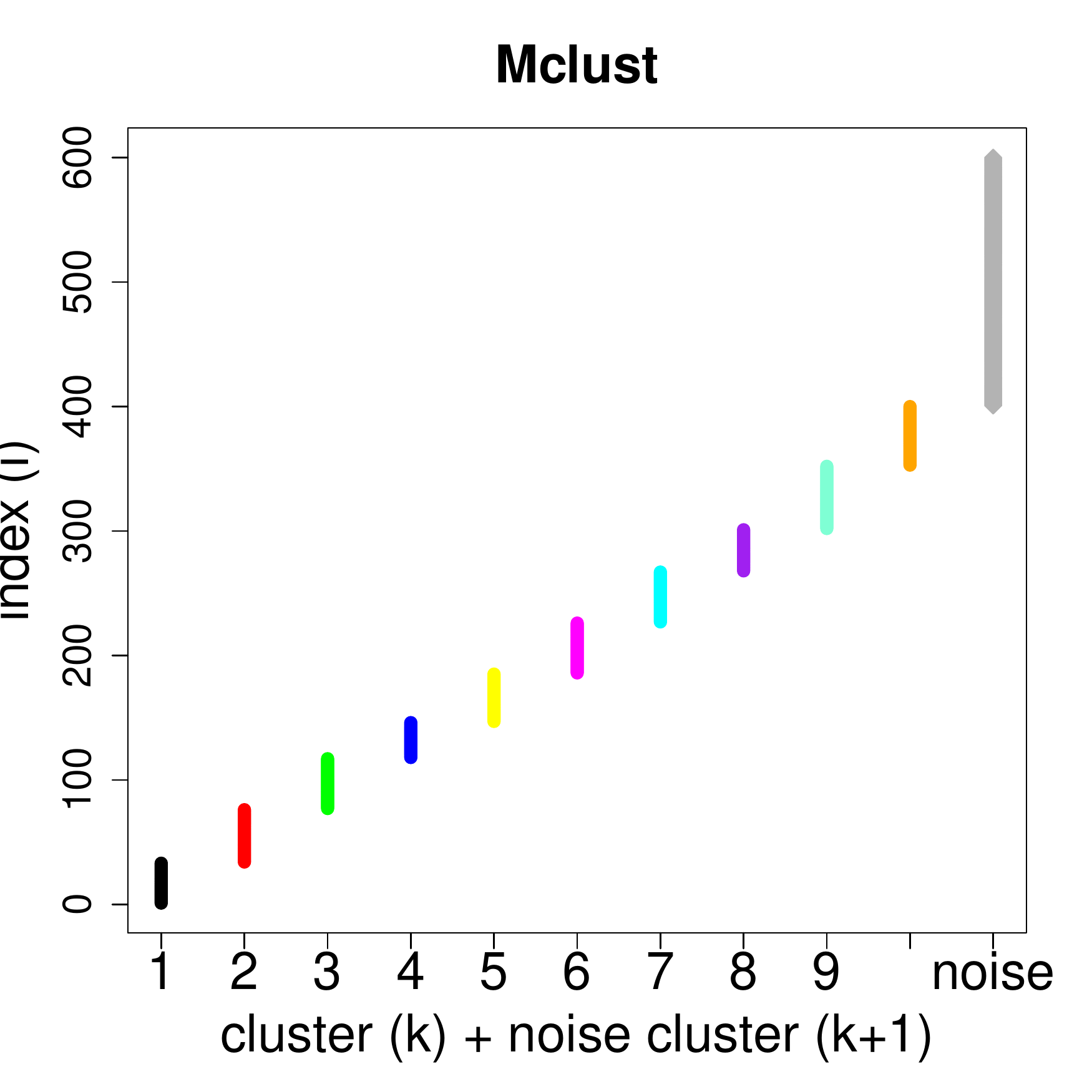}
		\caption{}
		\label{subfig44}
	\end{subfigure}
	\begin{subfigure}[b]{0.32\textwidth}
		\centering
		\includegraphics[width=\textwidth]{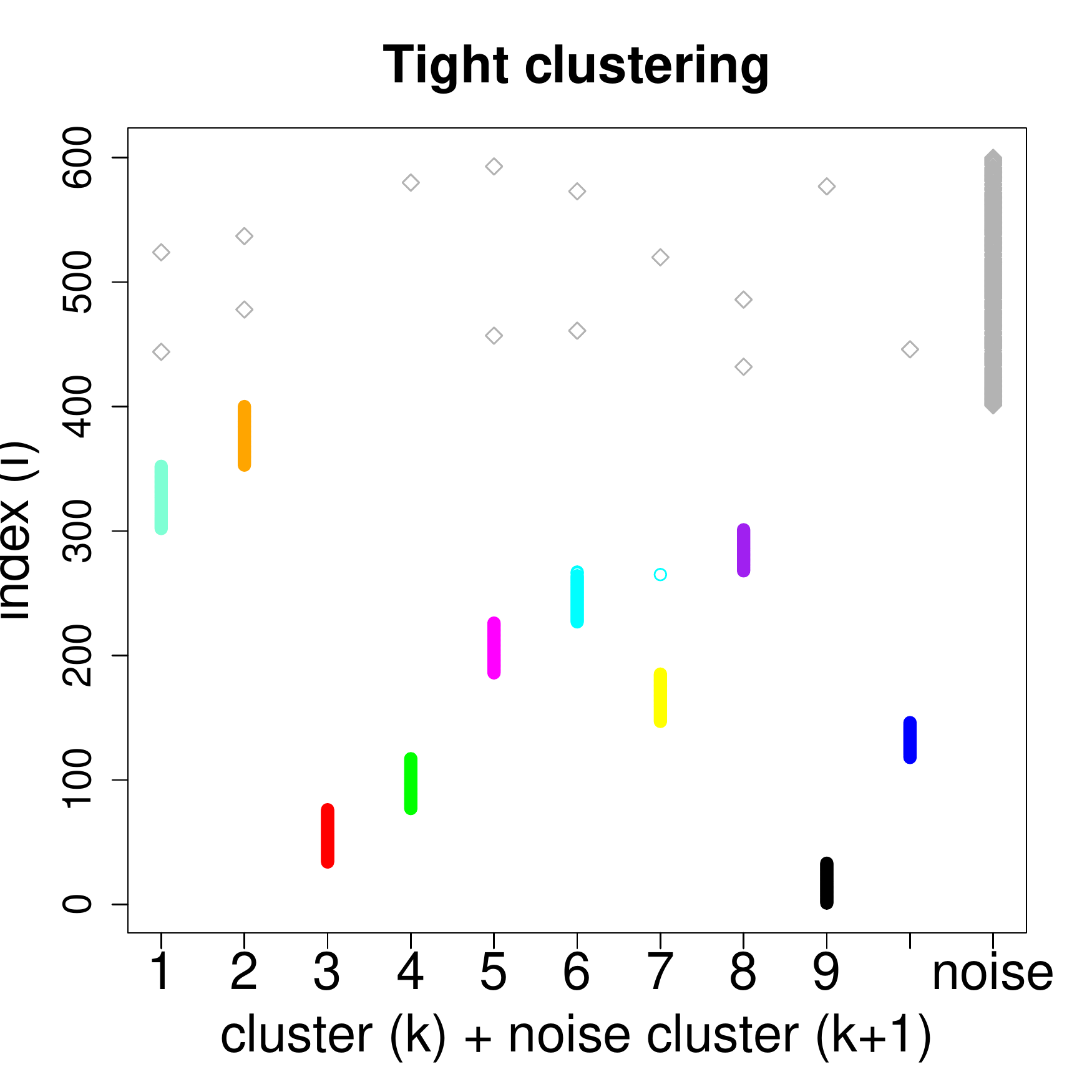}
		\caption{}
		\label{subfig45}
	\end{subfigure}
	\begin{subfigure}[b]{0.32\textwidth}
		\centering
		\includegraphics[width=\textwidth]{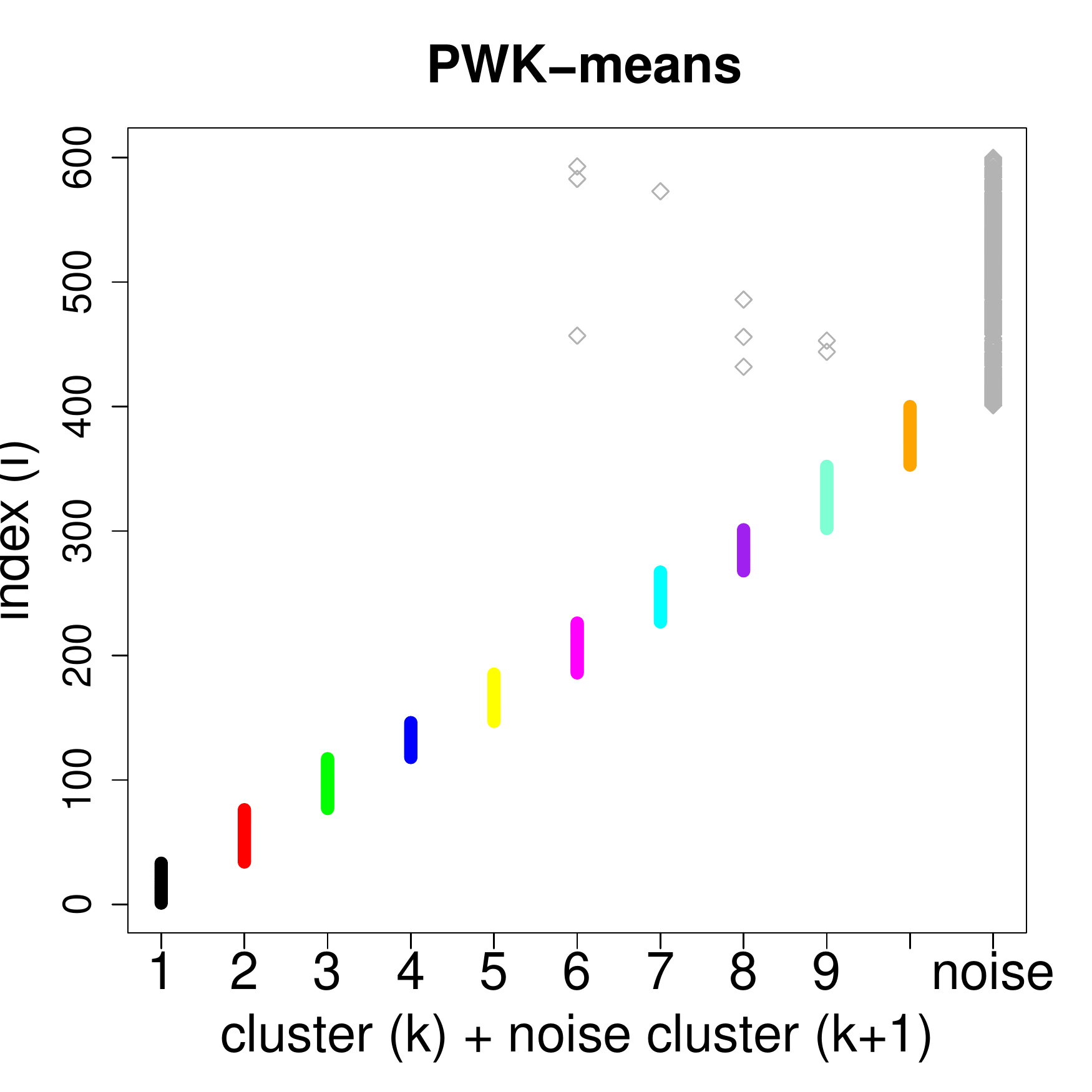}
		\caption{}
		\label{subfig46}
	\end{subfigure}
	\caption{Cluster assignment plots for comparison methods. The $y$-axis displays the index $i$ of each data point and the $x$-axis shows $K$ estimated clusters and one additional cluster for the noise. The color indicates the true cluster assignment with true noise in gray, such that if a clustered data point is identified as noise it will appear as noise on the $x$-axis but will have a color associated with it. If noise is misclassified then it will appear in the clusters but in gray color. (a)-(b) Cluster assignments for solution path clustering for $K=10$ and $K=9$, corresponding to two consecutive solutions in a solution path. (c)-(f) Cluster assignments for $K=10$ for convex clustering (CVX), mclust, tight clustering, and PWK-means.}
	\label{fig4}
\end{figure}

Figure \ref{fig4} shows an example of the cluster assignment for the overlapping scenario with noise for different methods. Figures \ref{subfig41}-\ref{subfig42} show cluster assignments for two consecutive solutions with $K=10$ and $K=9$ for SPC. The solution for $K=10$ leaves out a number of clustered data points as noise. The solution with $K=9$ separates the noise perfectly, but merges two overlapping clusters. Figures~\ref{subfig43}-\ref{subfig46} show the results of competing methods with $K=10$. One sees that mclust (Figure~\ref{subfig44}) separates noise well and misclassifies only one noisy point, while tight clustering (Figure \ref{subfig45}) and PWK-means (Figure \ref{subfig46}) add noise to the clusters. Convex clustering (Figure~\ref{subfig43}) has the largest number of noise added to the clusters. 

\begin{figure}[ht]
	\centering
	\begin{subfigure}[b]{0.49\textwidth}
		\centering
		\includegraphics[width=\textwidth]{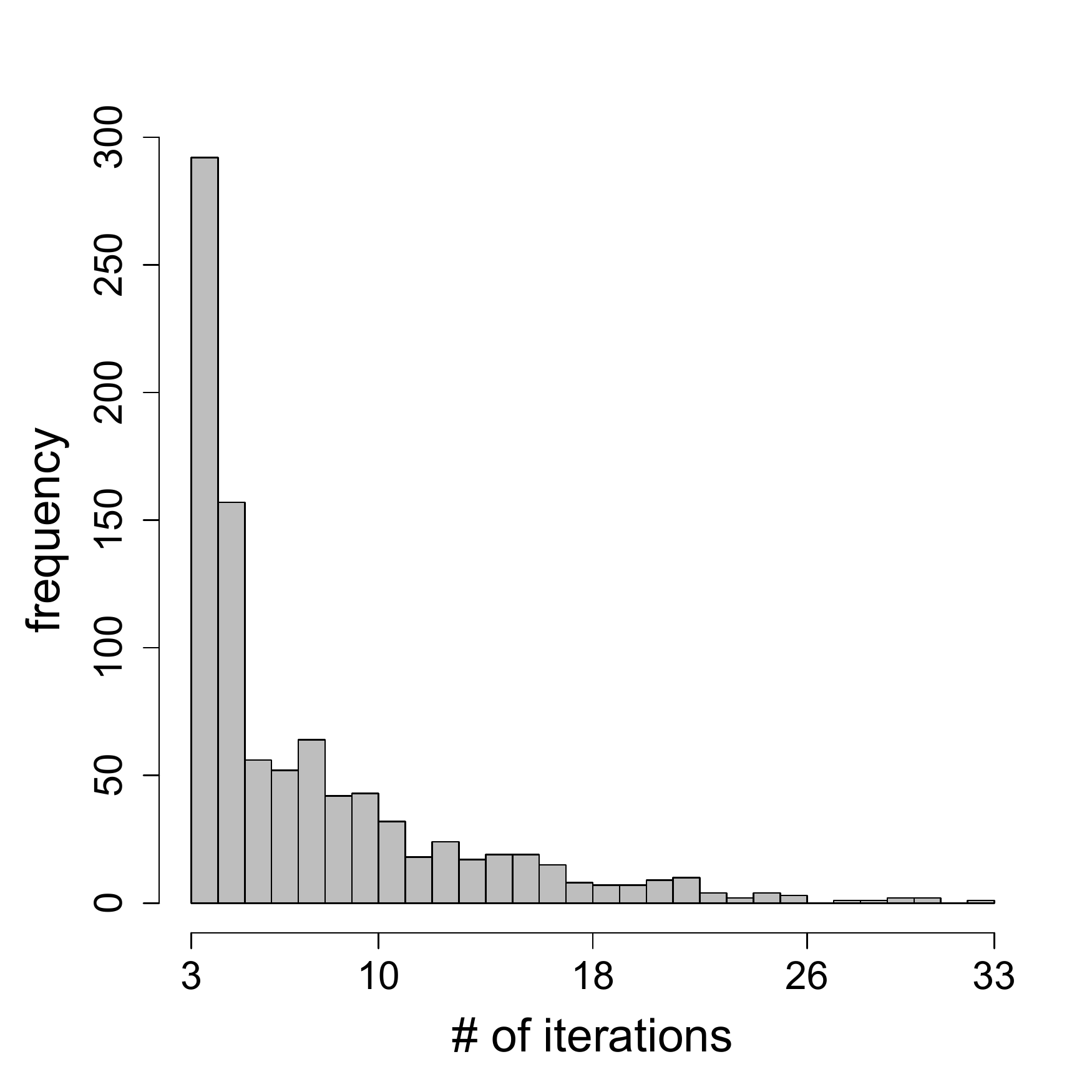}
		\caption{$n>p$}
		\label{it_bign}
	\end{subfigure} %
	\begin{subfigure}[b]{0.49\textwidth}
		\centering
		\includegraphics[width=\textwidth]{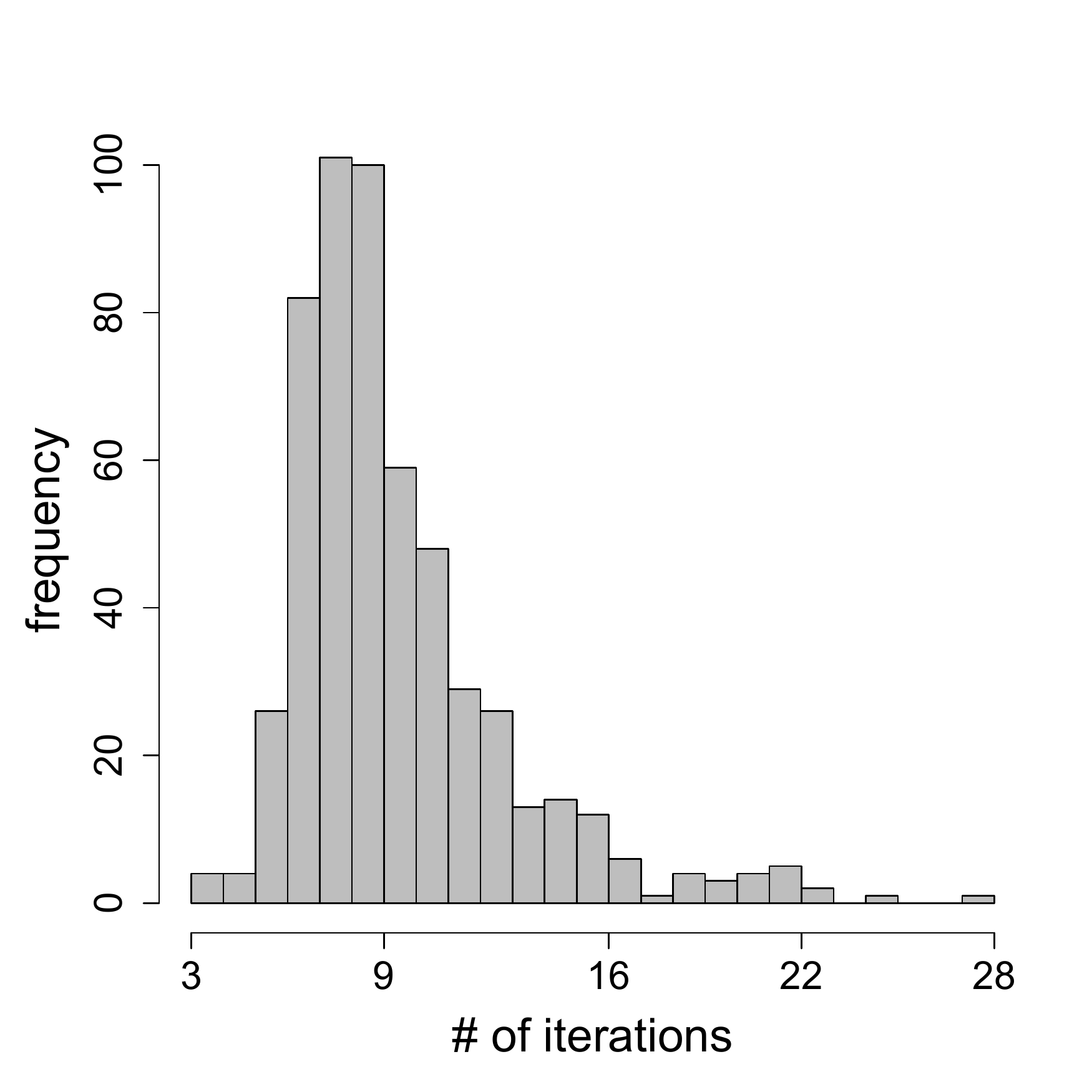}
		\caption{$n<p$}
		\label{it_bigp}
	\end{subfigure} %
	\caption{The number of iterations per solution for the combined four scenarios for the simulated data.}
	\label{it_hist}
\end{figure}

Finally, Figure~\ref{it_bign} shows the histogram of the number of iterations per solution for the combined four scenarios. Empirically, there were no instances of the algorithm not converging and the number of iterations per solution was between 3 to 33. The number of iterations to convergence generally decreases along a solution path because of warm starts and the reduction in $K$. When the solution gave the true number of clusters or was close to that, the algorithm converged within about 10 iterations. 

\subsection{Results for $n<p$}
\label{pn}

For $n<p$ we simulated datasets with $n=100$ clustered data points, $p=200$, and $K=10$. For each dataset, $50$ uniformly distributed noise points were added to the clustered data. Similarly to the scenarios with $n>p$, we consider clusters of size $N_k \leq 3$ as noise so that the results can be compared across the methods.  

\begin{figure}
	\centering
	\includegraphics[width=\textwidth]{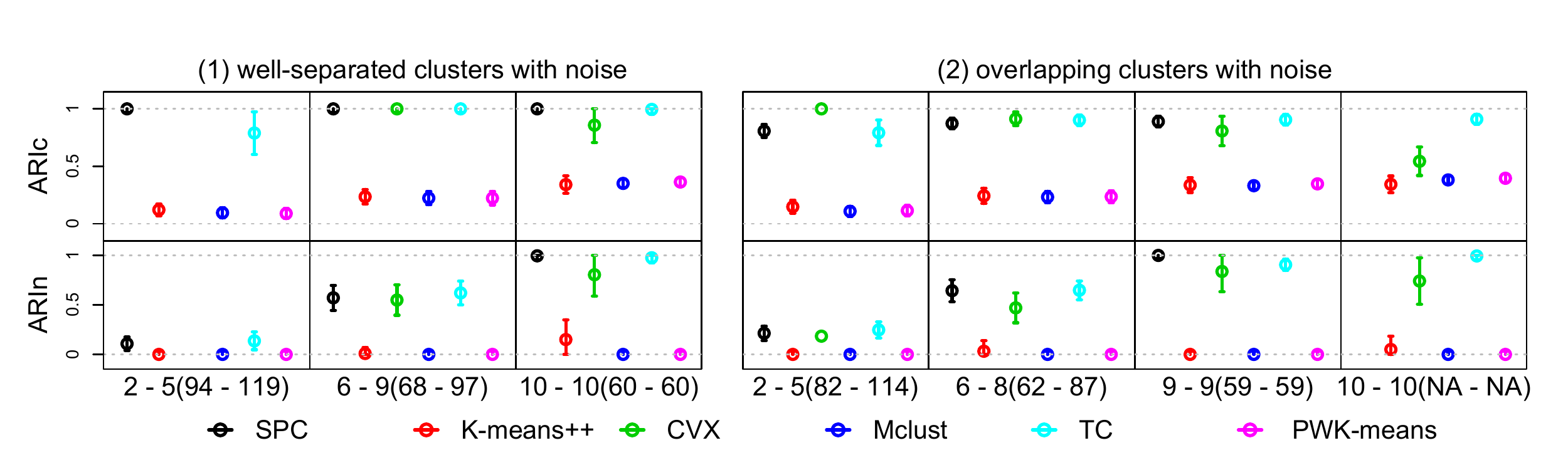}
	\caption{$\text{ARI}_c$ and $\text{ARI}_n$ for comparison methods for two scenarios with noise when $n<p$. The $x$-axis indicates $K_{\text{clust}}$ ranges and $K_{\text{total}}$ ranges in brackets. NA on the $x$-axis  in the last column in scenario (2) reflects that SPC did not obtain solutions with $K=10$ for this scenario.}
	\label{ARI_bigp}
\end{figure}

With high-dimensional data we chose $\omega = 0.1$ in order to create a longer and a more detailed solution path.  By setting a smaller $\omega$, fewer observations merge into clusters in the initial stages so that the majority of data points are left as singletons or very small clusters. As the sparsity is increased, more observations form new clusters. Consequently, even though the total number of clusters ($K_{\text{total}}$) decreases, the number of clusters of size $N_k > 3$ ($K_{\text{clust}}$) increases along the solution path. As shown in the results for SPC in Figure~\ref{ARI_bigp}, $\text{ARI}_c$ is always high because almost no noisy points merge into clusters and there are few misclassified clustered data points. On the other hand, $\text{ARI}_n$ increases as more clusters are formed so that fewer data points are regarded as noise. One sees that when $K_{\text{clust}}$ is close to the true number of clusters ($K=10$), satisfactory results are obtained in terms of both cluster assignment and noise detection.

When the number of dimensions $p$ is higher than the number of observations $n$, SPC shows a better performance than all the other methods in most scenarios. It performs at least as well as or slightly better than tight clustering. We could not obtain solutions for mclust due to error messages when the initial noise categorization was provided for the EM initialization and, thus, we used the solutions without the pre-specification of noise, which resulted in all the noise included in the clusters. This clearly demonstrates that mclust is sensitive to pre-specification of noise. K-means++ recognized clustered data quite well but grouped the noise with the clusters. The performance of mclust and k-means++ was very similar to that of PWK-means, which also did not recognize noise with penalty parameter $\lambda$ selected by prediction-based resampling. This comparison shows that identification of noise in high-dimensional space is very challenging and the uniform assumption for the noise may not be appropriate because the volume of the data range becomes huge when $p$ is large. 

Convex clustering showed slightly worse ARI scores in the $n<p$ scenarios, adding noise to the clusters as well as leaving out the clustered data as noise, which indicates that high dimensionality might create challenges for this method. The distances between observations in high-dimensional data become very large, forcing the parameter $\phi$ to be very close to 0 in order to produce reasonably large weights $w_{ij}$, which corresponds to uniform $w_{ij}$ and introduces more bias into the estimation of the cluster centers, possibly leading to unsatisfactory solutions. The $k$-nearest-neighbor approach could mitigate this problem somewhat but, as mentioned in qualitative comparisons section in \citet{chi2013}, can still lead to data points not being agglomerated correctly.  

As in the previous section, we demonstrate that the algorithm converged in all the scenarios for $n<p$ in Figure~\ref{it_bigp}. The number of iterations were somewhere between 3 and 28 for each solution. In the high-dimensional case it took on average more iterations (about 8 iterations vs. about 4 for $n>p$) to converge when the solution was at the true number of clusters or close.

Table~\ref{table:runtime} summarizes an example of the running time for all the methods except kmeans++, which is by far the fastest method. The run time for SPC is based on the longest solution path for each scenario, usually 10-14 solutions. The matching run times for mclust, tight clustering and PWK-means are calculated based on the unique cluster counts and those for convex clustering are based on its corresponding full solution path of length 20. If the number of clusters $K_{\text{clust}}$ (of size $N_k > 3$) is duplicated in a solution path, these duplicates are not included in the run times of mclust, tight clustering and PWK-means. The running time for PWK-means includes penalty parameter search with prediction-based resampling. SPC is currently implemented in R, and tight clustering and PWK-means are written in C, mclust in Fortran, and convex clustering in R and Fortran. It seems that SPC's run times compare well to the other methods and that it has a good potential in terms of speed, especially after implementation in a faster language.

\begin{table}[ht]
\begin{center}
\begin{footnotesize}
\caption{Summary of run times (in seconds) for SPC and comparison methods}
\label{table:runtime}
\begin{footnotesize}
\begin{threeparttable}
\begin{tabular*}{1\textwidth}{@{\extracolsep{\fill}}r|ccccc|ccccc}
  \hline
\multicolumn{1}{@{}c@{}|}{ } & \multicolumn{5}{@{}c@{}|}{ $n>p$ }  &  \multicolumn{5}{@{}c@{}}{ $n<p$ }  \\ 
 \multicolumn{1}{c@{}|}{} &  \multicolumn{1}{c}{SPC} & \multicolumn{1}{c}{CVX} & \multicolumn{1}{c}{mclust} & \multicolumn{1}{c}{TC} & \multicolumn{1}{c|}{PWK } &  \multicolumn{1}{c}{SPC} & \multicolumn{1}{c}{CVX} & \multicolumn{1}{c}{mclust} & \multicolumn{1}{c}{TC} & \multicolumn{1}{c}{PWK } \\ 
\hline
 (1) & 4.56 & 1.87 & 1.36 & 9.42 & 19.07  & 4.26 & 3.63 & 0.92 & 812.75 & 17.60 \\ 
  (2) & 3.88 & 3.76 & 1.34 & 12.61 & 19.26 & 2.68 & 3.36 & 0.95 & 369.51 &  14.29  \\
  (3) & 11.95 & 19.14 & 1.43 & 10.32 & 39.40 & 7.38 & 1.69 & 0.89 & 11.16 & 4.29  \\ 
  (4) & 11.43 & 24.50 & 1.56 & 10.36 & 41.29 & 9.10 & 0.76 & 1.08 & 11.92 & 4.64 \\ 
   \hline
\end{tabular*}
\begin{tablenotes}
\item NOTE: (1) well-separated clusters, (2) overlapping clusters, (3) well-separated clusters with noise, (4) overlapping clusters with noise. 
\end{tablenotes}
\end{threeparttable}
\end{footnotesize}
\end{footnotesize}
\end{center}
\end{table}

\subsection{Results for non-convex and non-spherical clusters}
\label{nonc}

\begin{figure}
	\centering
	\includegraphics[width=\textwidth]{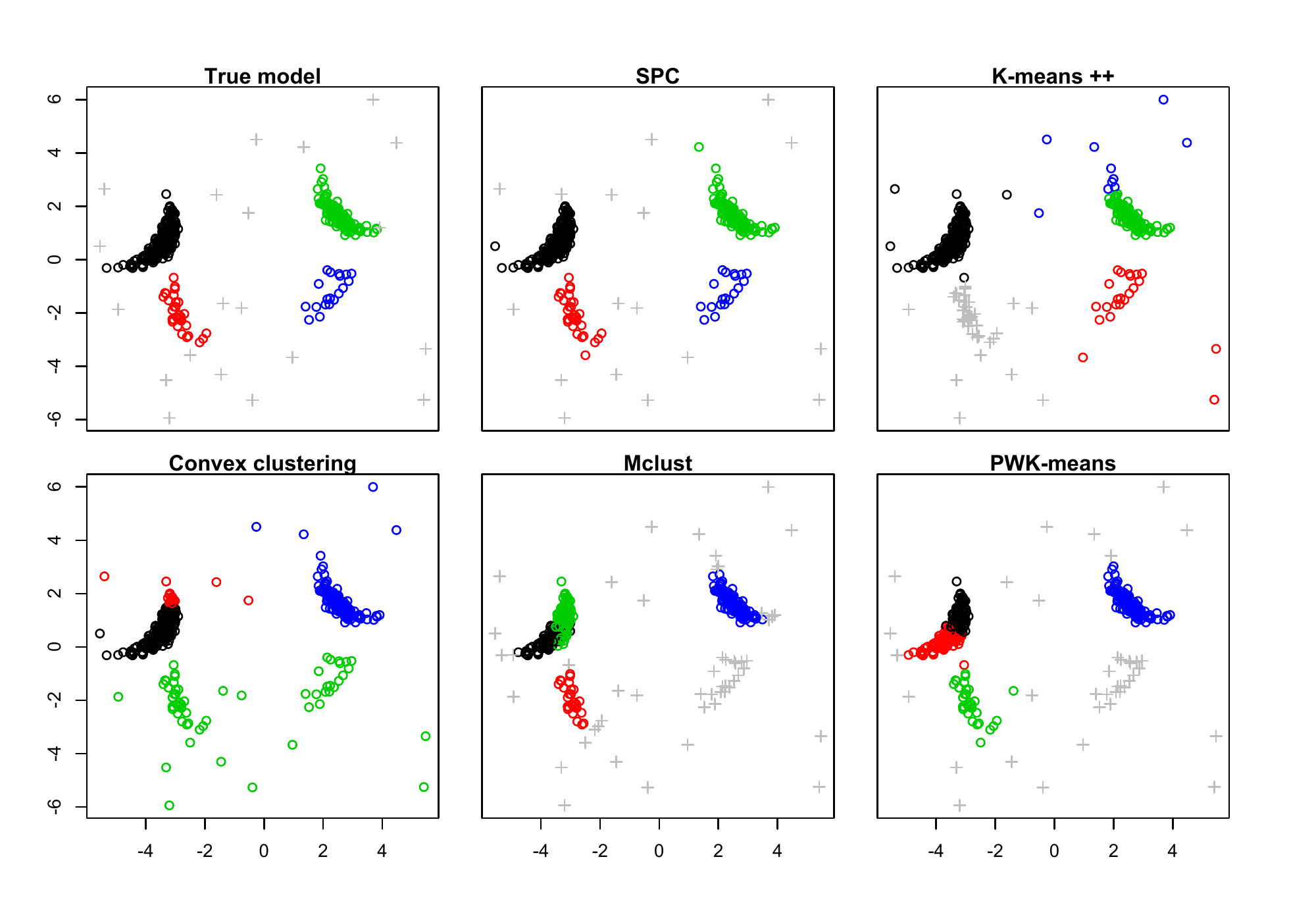}
	\caption{Cluster assignment results for non-convex clusters. Colored points indicate data in clusters and gray points indicate noise. }
	\label{nonconvex}
\end{figure}

We now show the performance of SPC on non-convex and non-spherical simulated clusters. For the non-convex case we generated $n=400$ observations in $p=2$ dimensions grouped in $K=4$ clusters, where data in each cluster were generated from a normal distribution with a high correlation. Three out of the four clusters have a negative or positive correlation of around 0.9 and one has a slightly lower correlation of 0.5. After the data points were simulated, non-convexity was introduced to each cluster as shown in Figure~\ref{nonconvex} (true model). The cluster size was varied to be $N_1 \approx 250$, $N_2 \approx 100$, $N_3 \approx 30$, and $N_4 \approx 20$. Finally, $20$ noise data points were added similarly to the scenarios in the previous sections. Such situation would be typical, for example, for gene expression data. We then applied all the methods except tight clustering, which did not converge, to this simulated data. 

It can be seen from Figure~\ref{nonconvex} that SPC can perform well in relatively complex settings when the clusters are non-convex, the noise or outliers are present and when the cluster sizes are different. All the comparison methods tend to split one of the two bigger clusters (black and green in the true model plot), while SPC does not. Mclust and PWK-means perform well with noise, however, they do not recognize the smallest cluster with the lowest correlation (blue). Mclust also assigns the ends of the non-convex clusters to noise, showing its sensitivity to the assumption of normality. Convex clustering does not seem to be robust to outliers and noise, due to the design of its penalty, specifically the weights, even though it can obtain very good results for non-noisy non-convex data for exactly the same reason.

\begin{figure}
	\centering
	\includegraphics[width=0.7\textwidth]{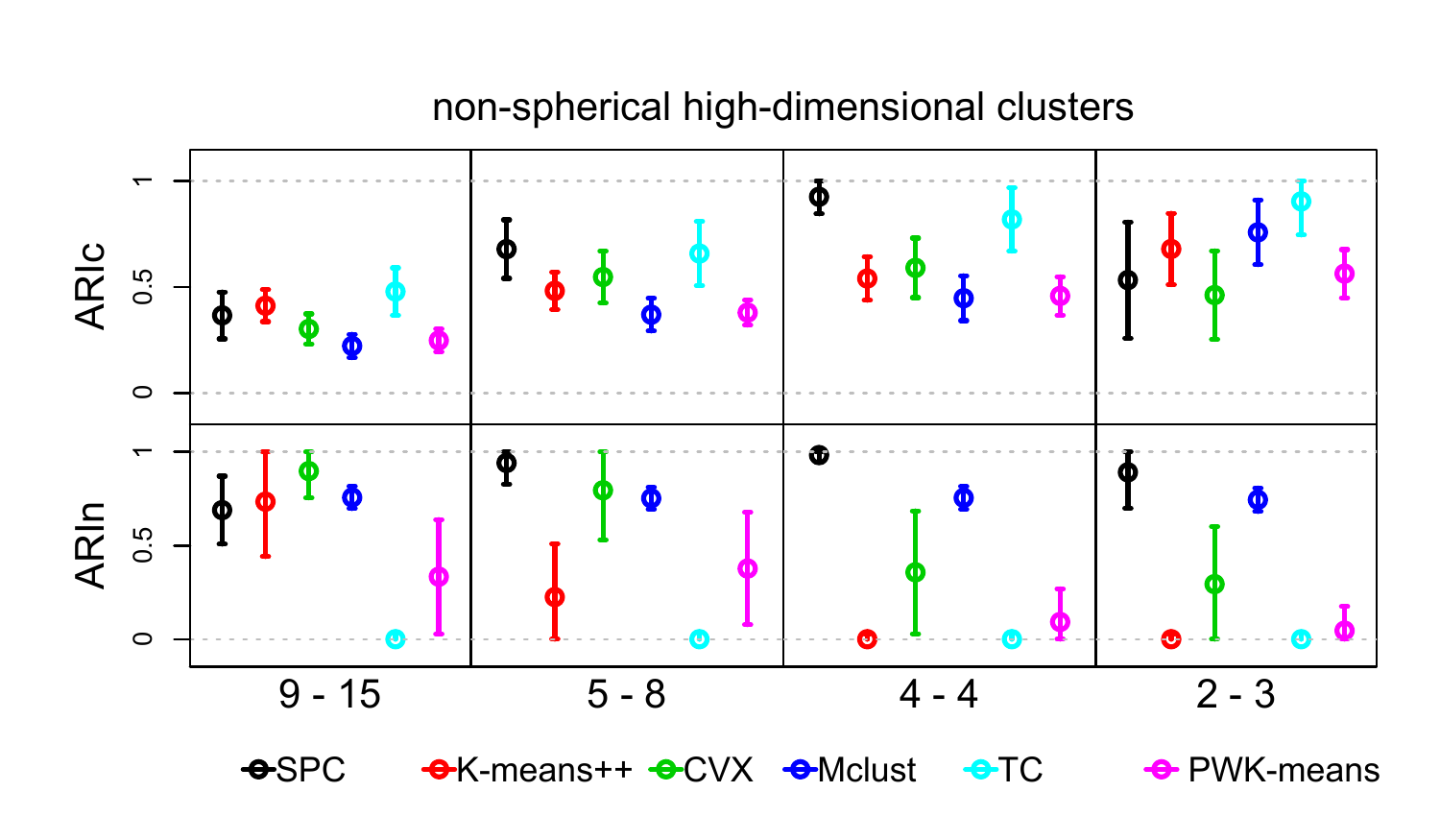}
	\caption{$\text{ARI}_c$ and $\text{ARI}_n$ for comparison methods for non-spherical simulated clusters with $p=20$ and $K=4$ true number of clusters.}
	\label{nonspherical}
\end{figure}

To further test the relative performance between SPC and the competing methods we applied them in higher dimensions. We generated similarly correlated $K=4$ clusters in $p=20$ dimensions with the same sizes and added $50$ noise points. We did not, however, introduce any non-convexity into the clusters. The resulting ARI scores, averaged over 20 randomly generated datasets, are presented in Figure~\ref{nonspherical}. SPC has the highest $\text{ARI}_c$ and $\text{ARI}_n$ scores for the true number of clusters $K=4$ and also performs well when the number of clusters is mis-specified. This confirms the usefulness of SPC for clustering high-dimensional non-spherical data.

\subsection{Sensitivity to tuning parameters}
\label{sensitivity}

In this section, we further comment on the sensitivity of the solution path to the tuning parameters $\omega$, $\phi$, $\tau$ in \eqref{eq:lambda1_1}, and $\alpha$ in \eqref{eq:delta}. As mentioned in the previous section, the parameter $\omega$ could have a considerable impact by skipping the best solution in the initial step if set too high. This can be avoided by setting this parameter to a low value at the expense of a more detailed solution path with a longer running time. We have found, however, that the rest of the tuning parameters do not impact the solution path in any major way. Setting a different value for $\alpha$ will automatically trigger a corresponding change in the value of $\lambda$, which might affect the speed of convergence. To further demonstrate the sensitivity to $\tau$ and $\phi$, we have run SPC on 20 randomly generated datasets with different combinations of these parameters and averaged the length of the solution path with $\text{ARI}_c$ and $\text{ARI}_n$ for each combination. Table \ref{table:sensitivity} is based on the overlapping case with noise for $n>p$ and illustrates that the cluster assignment  and noise detection are not sensitive at all to these two parameters. The length of the solution path, except for the smallest values for both $\tau$ and $\phi$, is also very stable. The greatest difference between the smaller and larger values of these tuning parameters was the speed of each solution due to the different minimization step size.

\begin{table}[ht]
\begin{center}
\caption{Solution path length and ($\text{ARI}_c$,  $\text{ARI}_n$) scores for different tuning parameters $\phi$ and $\tau$}
\begin{footnotesize}
\begin{tabular*}{1\textwidth}{@{\extracolsep{\fill}}r|@{\hspace{0.15cm}}l@{\hspace{0.15cm}}l@{\hspace{0.15cm}}l@{\hspace{0.15cm}}l@{\hspace{0.15cm}}l}
  \hline
  \multicolumn{1}{l|@{\hspace{0.15cm}}}{$\tau$} & \multicolumn{1}{l}{$\phi=0.05$} & \multicolumn{1}{l}{$\phi=0.1$} & \multicolumn{1}{l}{$\phi=0.3$} & \multicolumn{1}{l}{$\phi=0.5$} & \multicolumn{1}{l}{$\phi=0.7$}  \\ 
 \hline
  0.05 & 6 \, (0.91, 0.99) & 9 \,  (0.90, 1.00) & 10  (0.91, 1.00) & 10  (0.91, 1.00) & 11  (0.91, 0.99) \\ 
  0.15 & 8 \, (0.90, 1.00) & 13  (0.90, 1.00) & 10  (0.91, 1.00) & 11  (0.91, 0.99) & 11  (0.91, 0.99)  \\ 
  0.25 & 10  (0.90, 1.00) & 10  (0.90, 1.00) & 10  (0.91, 1.00) & 11  (0.91, 0.99) & 11  (0.92, 0.99)  \\ 
  0.35 & 11  (0.90, 1.00) & 10  (0.91, 1.00) & 11  (0.91, 0.99) & 11  (0.91, 1.00) & 11  (0.91, 0.99)  \\ 
  0.45 & 10  (0.91, 0.99) & 10  (0.91, 1.00) & 11  (0.91, 1.00) & 11  (0.92, 0.99) & 11  (0.92, 0.99) \\ 
\hline
\end{tabular*}
\end{footnotesize}
\label{table:sensitivity}
\end{center}
\end{table}

\section{Solution selection}
\label{solselection}

SPC does not require the specification of the number of clusters, however, it produces a solution path. Unlike the entire clustering path of hierarchical clustering or arbitrary regularization paths of convex clustering \citep{chi2013,hocking2011} or PRclust \citep{pan2013}, a SPC solution path includes only a limited number of solutions obtained using adaptive data-driven approach. This makes it easier for a user to explore different possible clustering assignments. However, it is still very useful in practice to be able to select a particular member along the solution path, especially for large datasets.  

Resampling-based methods such as, for example, the gap statistic \citep{tibshirani2001} or the clustering instability method in \cite{fang2012}, are computationally intensive, while Bayesian information criterion and cross-validation have been shown to select models that are too complex compared to the true model in sparse linear regression. These problems are especially relevant for our method that aims at high-dimensional and large data. We, therefore, adopt the empirical approach in \citet{fu2013} to demonstrate that solution selection for SPC is possible in a simple and fast way.  

\begin{figure}[ht]
	\centering
	\begin{subfigure}[b]{0.49\textwidth}
		\centering
		\includegraphics[width=\textwidth]{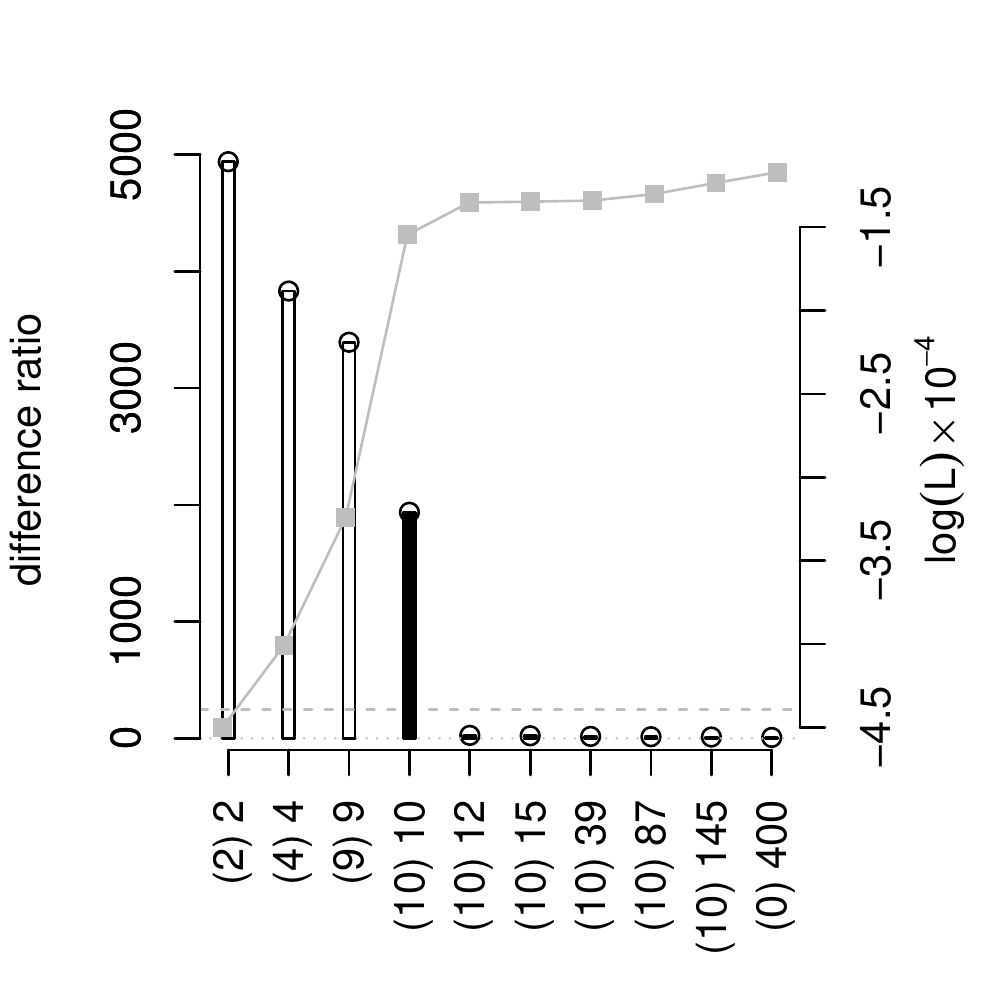}
		\caption{}
		\label{select1}
	\end{subfigure} %
	\begin{subfigure}[b]{0.49\textwidth}
		\centering
		\includegraphics[width=\textwidth]{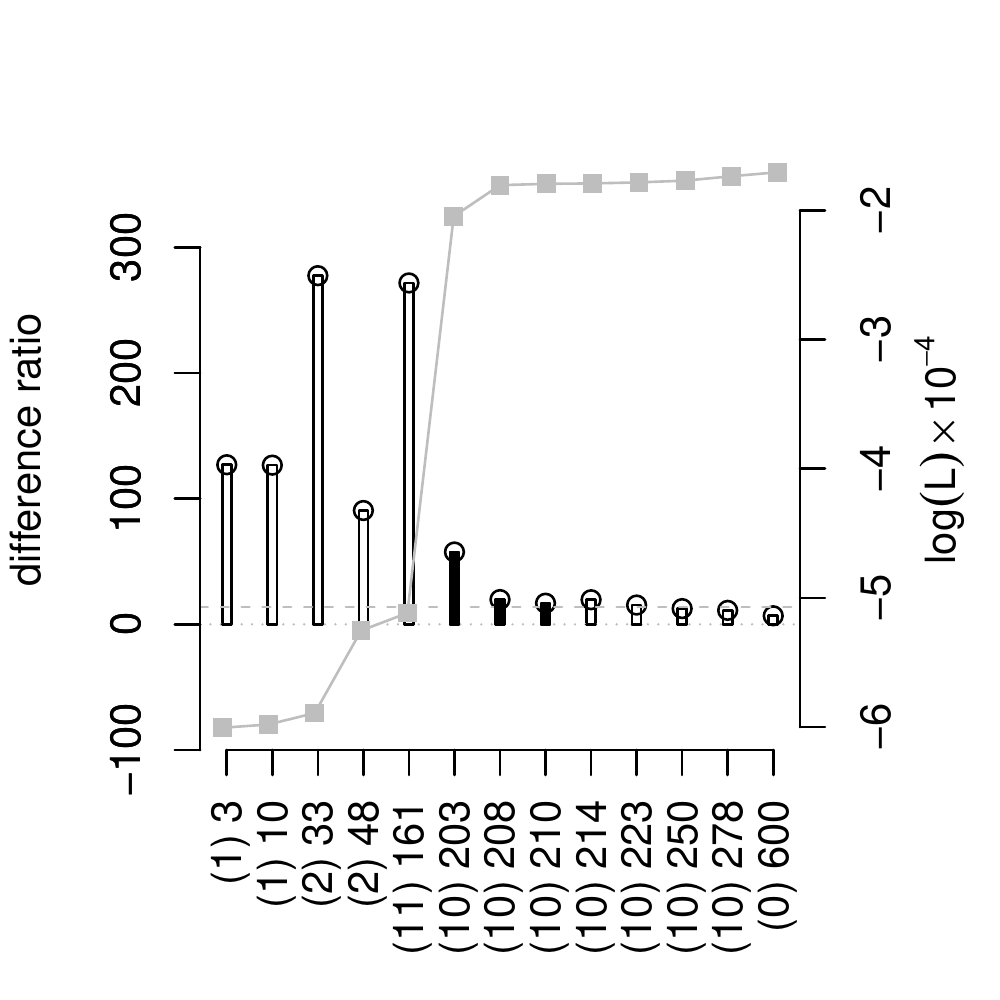}
		\caption{}
		\label{select2}
	\end{subfigure} %
	\caption{The difference ratio (vertical bars) and the unpenalized log-likelihood (gray solid line) as a function of the number of clusters in a solution path. The dashed line represents the cutoff with $a=0.05$ and the black color highlights the difference ratio for the solutions with the highest ARI scores. The number of clusters on the $x$-axis is represented by the total number of clusters $K_{\text{total}}$ and the number of clusters of size $N_k > 3$ in brackets ($K_{\text{clust}}$). (a) An example from the $n>p$ scenario with well-separated clusters and no noise. (b) An example from the $n>p$ scenario with well-separated clusters and noise.}
	\label{selection}
\end{figure}

The approach of \citet{fu2013} is based on the fact that as sparsity decreases, i.e. the number of clusters increases, the unpenalized log-likelihood of the data will increase. The increase in the number of clusters, then, is justifiable only by a significant increase in the unpenalized log-likelihood. We should choose a solution after which an increase in the number of clusters will not correspond to a big increase in the unpenalized log-likelihood. To determine this, we sort the solution path according to an increasing number of clusters  $K_{\text{total}}$ and, for each solution $s$, $s = 1, \ldots, S$, we calculate the difference ratio for two adjacent solutions:
\begin{align}
	\label{eq:dr}
	& dr^{(s,s+1)} = \frac{ L(K^{(s+1)}) - L(K^{(s)}) }{  K^{(s+1)} - K^{(s)}  },
\end{align}
provided that $K^{(s+1)} - K^{(s)}\geq 1$, where
\begin{align}
	\label{eq:L}
	& L( K^{(s)} ) = \sum_{i=1}^n \log \left[  \sum_{k=1}^{K^{(s)}} \pi_k^{(s)} \phi( y_i; \mu_k^{(s)}, \Sigma_k^{(s)} ) \right]
\end{align}
is the mixture Gaussian log-likelihood function and $K^{(s)}$ is the total estimated number of clusters $K_{\text{total}}$ for that solution. The cluster centers $\mu_k^{(s)}$ and the cluster proportions $\pi_k^{(s)}$ are estimated given the cluster assignments for each solution $s$. We set the covariance to be the identity matrix $\Sigma_k^{(s)} = I_k^{(s)}$, in line with the implicit assumption behind the use of $\ell_2$ loss in \eqref{eq:genloss} and \eqref{eq:kloss}. The likelihood for a singleton cluster in this case is well-defined and in effect amounts to $1/n$. We choose the solution indexed by
\begin{align}
	\label{eq:K}
	& K^* = \max \left\{ K^{(s)}: dr^{(s,s+1)} \geq a \times \max \left( dr^{(1,2)}, \ldots, dr^{(S-1, S)} \right) \right\},
\end{align}
with $a=0.05$ as suggested by \citet{fu2013}. We have plotted the difference ratio \eqref{eq:dr} and the unpenalized log-likelihood \eqref{eq:L} for two simulated datasets in Figure~\ref{selection}. In Figure~\ref{select1} the selected solution according to \eqref{eq:K} is the solution with the true number of clusters $K^* = K_{\text{total}} = K_{\text{clust}}=10$. An example from the scenario with noise in Figure~\ref{select2} is more ambiguous and in this particular case we would barely choose a solution with $K_{\text{total}} = 214$ and $K_{\text{clust}}=10$, which does not correspond to the highest ARI score and leaves 4 of the clustered points as noise. Generally, however, the selected solutions have relatively high ARI scores, as reported in Table~\ref{table:selection}. In this table, we demonstrate the averages for $\text{ARI}_c$ and $\text{ARI}_n$ scores for the selected solutions as well as the averages for the best ARI scores (in terms of the sum of the two ARI scores) along each solution path of the corresponding scenario. One sees that the ARI scores of a selected solution are close to the best along the solution path. The table also reports the average selected $K_{\text{clust}}$ and its range, showing that the number of clusters given by a selected solution is close to the true one. All averages and ranges are taken over 20 simulated datasets.

\begin{table}[ht]
\begin{center}
\begin{footnotesize}
\caption{Solution selection summary for $n>p$ scenarios}
\label{runtime}
\begin{footnotesize}
\begin{threeparttable}
\begin{tabular*}{0.8\textwidth}{@{\extracolsep{\fill}}r|cccccc}
  \hline
\multicolumn{1}{c|}{ } & \multicolumn{2}{c}{$\text{ARI}_c$}  & \multicolumn{2}{c}{$\text{ARI}_n$} & \multicolumn{2}{c}{$K_{\text{clust}}$}  \\ 
 \multicolumn{1}{c|}{} &  \multicolumn{1}{c}{select} &  \multicolumn{1}{c}{best} &  \multicolumn{1}{c}{select} &  \multicolumn{1}{c}{best} &  \multicolumn{1}{c}{mean} &  \multicolumn{1}{c}{range}  \\
\hline
  (1) & 1.000 & 1.000 & 1.000 & 1.000 & 10 & 10-10 \\ 
  (2) & 0.899 & 0.935 & 1.000 & 0.986  & 9 & 9-9   \\
  (3) &  0.986 & 1.000 & 0.979 & 1.000 & 10 & 10-13  \\ 
  (4) & 0.940 & 0.918 & 0.900 & 0.987 & 10 & 9-10  \\ 
   \hline
\end{tabular*}
\begin{tablenotes}
\item NOTE: (1) well-separated clusters, (2) overlapping clusters, (3) well-separated clusters with noise, (4) overlapping clusters with noise. $\text{ARI}_c$ and $\text{ARI}_n$ are averaged over 20 datasets for each scenario for the selected solutions (select) and for the largest ARI scores in each solution path (best).
\end{tablenotes}
\end{threeparttable}
\label{table:selection}
\end{footnotesize}
\end{footnotesize}
\end{center}
\end{table}

It is generally hard to determine whether singleton or very small clusters are truly noise and, thus, solution selection is more challenging when noisy data points are present, which is reflected in selected average $\text{ARI}_n$ being slightly lower than the best $\text{ARI}_n$ scores in Table~\ref{table:selection}. For overlapping scenario with no noise, we tended to select the solutions with $K_{\text{clust}} = 9$ clusters while the best ARI scores were given by solutions with $K_{\text{clust}} = 10$ (with a few clustered data points misclassified as noise). On the contrary, for the overlapping scenario with noise, most of the solutions selected were given by $K_{\text{clust}} = 10$, which resulted in a higher than best $\text{ARI}_c$ but lower than best $\text{ARI}_n$. Altogether this method is very fast and appears to select good solutions for SPC, but more experiments are needed, especially for various possible assumptions on $\Sigma_k$.

\section{Clustering gene expression data}
\label{geneexp}

To further test the performance of SPC in a real data environment we applied it to a gene expression dataset \citep{zhou2007}. The full dataset consists of over 45,000 genes across 16 different experimental conditions that were created to study the regulation of these genes by Oct4, a transcription factor (TF) important for the self-renewal and maintenance of mouse embryonic stem cells. The 16 gene expression profiles include 3 generated from undifferentiated cells which are naturally high in Oct4 expression (conditions 1-3), 5 from early differentiated cells with high Oct4 expression (conditions 4-8), and 8 with low Oct4 expression (conditions 9-16). The study in \citet{zhou2007} has identified and referenced 1,325 Oct4-high genes (Oct4+) and 1,440 Oct4-low genes (Oct4$-$) out of all the genes, which we regard as two true separated clusters for validation. We then added 3,000 extra randomly selected genes with various levels of coefficient of variation and randomly permuted the expression vector of each gene to obtain noise data points. Thus, we analyzed a final dataset with $n=5,765$ observations (genes), $p=16$ dimensions, and $K=2$ distinct clusters of Oct4+ and Oct4$-$ genes. Following the common practice, we normalized the expression data of each gene to have zero mean and unit standard deviation.

\begin{figure}[ht]
	\centering
	\begin{subfigure}[b]{0.4\textwidth}
		\centering
		\includegraphics[width=\textwidth]{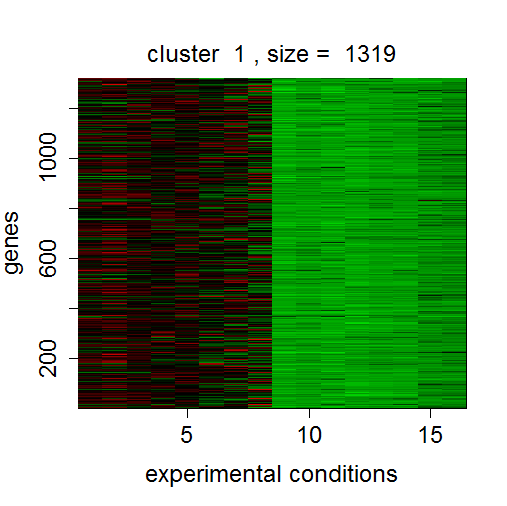}
		\caption{}
		\label{subfig51}
	\end{subfigure} %
	\begin{subfigure}[b]{0.4\textwidth}
		\centering
		\includegraphics[width=\textwidth]{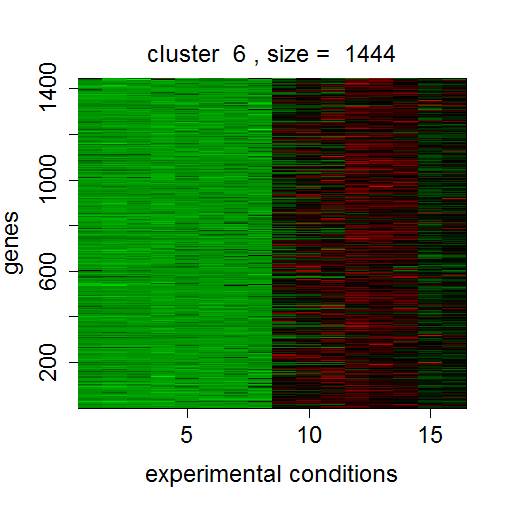}
		\caption{}
		\label{subfig52}
	\end{subfigure} %
	\begin{subfigure}[b]{0.4\textwidth}
		\centering
		\includegraphics[width=\textwidth]{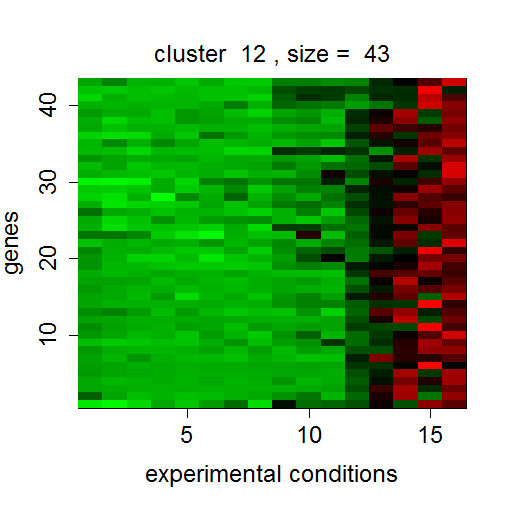}
		\caption{}
		\label{subfig53}
	\end{subfigure}
	\caption{Heatmaps of the three largest clusters identified by SPC. Red color indicates high expression and green color indicates low expression.}
	\label{fig5}
\end{figure}

We ran SPC with $\omega=0.5$ and the resulting solution path is presented in Table \ref{table:genedata}. SPC has clearly identified the largest two clusters of Oct4+ and Oct4$-$ genes which are shown in Figure \ref{subfig51}-\ref{subfig52}. It also suggested an additional smaller cluster of size $N_k = 43$ shown in Figure \ref{subfig53}.  We do not address in this paper the problem of choosing a proper threshold for the size of a cluster. For demonstration purposes we consider only $K=2$ largest clusters and regard the rest of the observations as noise, including the small cluster in Figure \ref{subfig53}. The left side of Table \ref{table:genedata} shows the number of clusters of different size per each solution and the right side shows the numbers of Oct4+, Oct4$-$, and random genes in the two largest clusters (Cluster 1 and Cluster 2) and those in the rest of the genes (Other). The differences among the solutions 1-4 are only due to the random genes forming mostly very small clusters of size $N_k < 30$, and there are only 4 random genes assigned to the Oct4 clusters.  It can be seen from the last column that a small number of Oct4 genes were left out as random, but most of these genes (43 out of 48 Oct4$-$ genes) formed the small Oct4$-$ cluster of Figure \ref{subfig53}.  In solution 5 the small  Oct4$-$ cluster is merged with the main Oct4$-$ cluster and another big cluster of size $>$ 1,000 is formed from many random genes. In solution 6 the majority of the genes merge into a single cluster and in the subsequent solutions the remaining random genes (outliers) are gradually added to this cluster. All the solutions in Table \ref{table:genedata} converged in fewer than 41 iterations.

\begin{table}[ht]
\begin{footnotesize}
\begin{center}
\caption{Solution path for the gene expression data}
\begin{threeparttable}
\begin{tabular*}{1\textwidth}{@{\extracolsep{\fill}}@{}r|c@{\hspace{0.5em}}c@{\hspace{0.5em}}|c@{}c@{}c|c@{/}c@{/}c@{\hspace{2em}}c@{/}c@{/}c@{\hspace{2em}}c@{/}c@{/}c}
 \hline
 \multicolumn{1}{c|}{  } & \multicolumn{2}{c}{  } & \multicolumn{3}{@{}c@{}|}{\# of clusters of size} & \multicolumn{9}{c}{\# of genes (Oct4+ / Oct4$-$ / noise)} \\
 & \multicolumn{1}{c@{}}{ $\delta$  } & \multicolumn{1}{@{}c}{$\lambda$} & \multicolumn{1}{@{}c@{}}{$[ 1\text{K}, n)$ } & \multicolumn{1}{@{}c@{}}{$[ 30, 1\text{K})$} & \multicolumn{1}{@{}c@{}|}{$( 1, 30 )$ } & \multicolumn{3}{c@{\hspace{1.5em}}}{Cluster 1} & \multicolumn{3}{c@{\hspace{1.5em}}}{Cluster 2} & \multicolumn{3}{c@{\hspace{1.5em}}}{ Other } \\
  \hline
1 & 0.3 & 6.2 & 2 & $1^*$  & 2954 & 1313 & 0 & 3 & 0 & 1392  & 1 & 12 & 48 & 2996 \\ 
  2 & 0.3 & 6.7 & 2 & $1^*$  & 2868 & 1313 & 0 & 3 & 0 & 1392 & 1 & 12 & 48 & 2996 \\ 
  3 & 0.3 & 7.3 & 2 & $1^*$  & 2617 & 1313 & 0 & 3 & 0 & 1392 & 1 & 12 & 48 & 2996 \\
  4 & 0.3 & 8.0 & 2 & $1^*$  & 2000 & 1313 & 0 & 3 & 0 & 1392 & 1 & 12 & 48 & 2996 \\ 
  5 & 0.3 & 8.7 & $3$ & $1$  & 837 & 1315 & 0 & 4 & 0 & 1435 & 9 & 10 & 5 & 2987 \\ 
  6 & 0.3 & 9.5 & 1 & 0  & 234 & 1325 & 1440 & 2658 & 0 & 0 & 8 & 0 & 0 & 334 \\ 
  7 & 0.3 & 10.4 & 1 & 0  & 80 & 1325 & 1440 & 2910 & 0 & 0 & 3 & 0 & 0 & 87 \\ 
  8 & 0.3 & 11.3 & 1 & 0  & 46 & 1325 & 1440 & 2954 & 0 & 0 & 1 & 0 & 0 & 45 \\ 
  9 & 0.3 & 12.4 & 1 & 0  & 4 & 1325 & 1440 & 2996 & 0 & 0 & 1 &  0 & 0 & 3 \\ 
  10 & 0.3 & 13.5 & 1 & 0  & 0 & 1325 & 1440 & 3000  & 0 & 0 & 0  &  0 & 0 & 0 \\ 
 \hline
\end{tabular*}
\begin{tablenotes}
\item NOTE: The asterisks $*$ identify the small cluster of size $N_k = 43$ (Oct4$-$) in the solution path.
\end{tablenotes}
\end{threeparttable}
\label{table:genedata}
\end{center}
\end{footnotesize}
\end{table}

Table~\ref{table:genecomp} compares the result of SPC with that of mclust, tight clustering, and PWK-means. For best results, we ran tight clustering with $K=2$ and $k_0 = K + 1$ since any larger values of $k_0$ produced inferior results, with large amounts of Oct4 genes added to the random category. For mclust, we overrode the default value for the hypervolume $V$ and applied the calculation from Section 4.1. For the nearest neighbor cleaning $K=5$ was used (the resulting partitioning into Oct4 clusters and random genes was not sensitive for $K < 40$, while if $K \geq 40$ was used, no noise would be identified in this data). Finally, for PWK-means we did the penalty parameter $\lambda$ search as with the simulated data. However, we had to start the sequence with a value greater than 1, otherwise the chosen value of $\lambda$ would be less than 1 by prediction-based resampling and would result in most of the genes classified as random. Overall, we felt that all the other methods required more user fine tuning and re-running for this dataset than SPC. It should be noted that the variation in the results of tight clustering across different runs was negligible, while the results of the other three methods were deterministic.

\begin{table}[t]
\begin{footnotesize}
\begin{center}
\caption{Comparison of different clustering methods for the Oct4-sorted expression data}
\begin{tabular*}{0.8\textwidth}{@{\extracolsep{\fill}}r|c@{ / }c@{ / }c@{\hspace{2em}}c@{ / }c@{ / }c@{\hspace{2em}}c@{ / }c@{ / }c}
\hline
  & \multicolumn{9}{c}{ \# of genes (Oct4+ / Oct4$-$ / noise) } \\
 & \multicolumn{3}{c@{\hspace{2em}}}{ Cluster 1 } &   \multicolumn{3}{c@{\hspace{2em}}}{ Cluster 2 } &  \multicolumn{3}{c@{\hspace{2em}}}{ Other }  \\ 
  \hline
SPC & 1315 & 0 & 3 & 0 & 1435  & 9 & 10 & 5 & 2987  \\ 
mclust &  1302 & 0 & 4  & 0 & 1412 & 5 & 23 & 28 & 2991  \\ 
tight clust & 1323 & 0 & 108  &  0 & 1402 & 64  & 2 & 38 & 2828 \\ 
PWK & 1250 & 0 & 11 & 0 & 1375 & 13 & 75 & 65 & 2976 \\ 
   \hline
\end{tabular*}
\label{table:genecomp}
\end{center}
\end{footnotesize}
\end{table}

It can be seen from Table~\ref{table:genecomp} that none of the methods misclassified genes between the Oct4+ and Oct4$-$ sets in this well-separated data, however, generally SPC added the fewest random genes into the Oct4 clusters and left out fewer Oct4 genes as random. SPC (solution 5 in Table \ref{table:genedata}) performed better compared to tight clustering, which tended to add more random genes into Oct4 clusters, and PWK-means, which left out more Oct4 genes as random. SPC identified a larger number of Oct4 genes compared to mclust. Even when comparing mclust to solutions 1-4 of SPC, it can be seen that SPC's overestimation of the number of random genes is attributed to the small Oct4$-$ cluster in Figure \ref{subfig53} that can arguably be classified as a separate cluster. Indeed, compared with those genes in the big Oct4$-$ cluster in Figure \ref{subfig52}, the genes in Figure \ref{subfig53} 
have lower expression levels in conditions 9 to 12 and higher expression levels in conditions 13 to 16, corresponding to cells starting differentiation. 

Gene ontology (GO) term enrichment analysis further confirms the specific biological roles
of the genes in this small cluster. Using the full set of the 1,440 Oct4$-$ genes as the background set, we found
64 enriched GO terms from the process ontology with $P<1.7\times 10^{-4}$ and three from
the function ontology with $P<1\times 10^{-4}$, both having a false discovery rate (FDR) of less than $0.005\%$ as strong evidence that these genes are not clustered by chance. From Table \ref{table:GO} we see that this small cluster is highly enriched for genes involved in nerve development and organismal process and for TFs with sequence-specific DNA binding activity, compared to the corresponding proportions in the full Oct4$-$ set.
This analysis demonstrates a strong performance of SPC on a larger gene expression dataset and the possibility of discovering smaller, biologically meaningful clusters by further examining the solution path.
Moreover, this shows an example where a detailed solution path is useful even when strong prior knowledge for the number
of clusters is available. With the expected number of clusters $K=2$, all the other methods failed to discover
this small cluster. The results of mclust, tight clustering and PWK-means identified solely the two big clusters and only differed in how many random genes were added to the clusters and how many Oct4 genes were left out as random.

\begin{table}[ht]
\begin{footnotesize}
\begin{center}
\caption{Top GO terms enriched in the small Oct4$-$ cluster}
\begin{threeparttable}
\begin{tabular*}{1\textwidth}{@{\extracolsep{\fill}}m{6cm}m{1cm}m{1cm}m{1.9cm}}
\hline
  GO term & \% in cluster & \% in full set & P-value  \\ 
  \hline \\ [-0.6em]
nervous system development & 43.6 & 12.9 & $1.41 \times 10^{-6}$  \\ 
multicellular organismal process & 71.8 & 34.8 & $1.99 \times 10^{-6}$  \\
cranial nerve development & 10.3 & 0.3 & $2.27 \times 10^{-6}$ \\ 
\\
 sequence-specific DNA binding & 25.6 & 5.6 & $2.76 \times 10^{-5}$  \\ 
 sequence-specific DNA binding TF activity & 28.2 & 6.8 & $2.86 \times 10^{-5}$ \\
nucleic acid binding TF activity & 28.2 & 6.9 & $3.16 \times 10^{-5}$ \\ [0.2em]
    \hline
\end{tabular*}
\begin{tablenotes}
\item Tabulated are the top three GO terms from the process (top) and  the function (bottom) ontology.
\end{tablenotes}
\end{threeparttable}
\label{table:GO}
\end{center}
\end{footnotesize}
\end{table}

\section{Discussion}
\label{discussion}

SPC is a new general clustering method that provides a small set of solutions, each including a cluster assignment and a number of clusters, in a wide variety of situations including high-dimensional settings when model-based clustering, for example, may have difficulty in providing a solution. SPC can be applied to a dataset of different complexity, whether it is small with only a few outliers or large with a high proportion of irrelevant observations. This method searches for the best solution given a certain degree of sparsity penalty, which is determined adaptively by the data. The irrelevant observations are simultaneously identified as singletons or very small clusters. SPC does not require extensive fine tuning and has only one required input tuning parameter, which could have some impact on the resulting set of solutions. However, even this parameter can be set at the lowest possible value, with the only drawback of a longer solution path and running time. Most importantly, SPC does not require the knowledge of the number of clusters and provides its own estimate, depending on the amount of sparsity imposed. 

We have assumed that once two or more objects merge they are not split in the later stages of the solution path, which is a shortcut that breaks down the direct optimization of the objective function \eqref{eq:genloss}. It is possible, however, to eliminate this assumption and perform the splitting via a soft thresholding operation for those $\theta_i = \theta_j$, as discussed in the Appendix. In practice we found that this step slows down the algorithm. It might be worthwhile to investigate whether there are cases when the splitting enables finding better solutions. Additionally, it might be interesting to apply soft thresholding when the SPC algorithm is initialized assuming all objects belong to one cluster and then gradually split the objects, similar to divisive hierarchical clustering.

Solution path clustering is effective in separating noise from clustered data, however, it tends to merge overlapping clusters due to its dependence on distances between cluster centers. In the later stages of a solution path, when sparsity is higher, it also tends to add more noise into clusters as the larger distances, usually characterized by noisy observations or outliers, receive more penalization. Overall, SPC have provided satisfactory clustering results across all the simulated data scenarios and the gene expression data. 

Penalization methods have been widely applied in linear model analysis and are emerging in clustering. The idea of a solution path or surface along with model selection methodology has been successfully utilized for fast search of sparse models.  We extended the regularization path or surface and sparsity approach to unsupervised learning, where the gradual increase of sparsity is guided by the data and the bias of the cluster centers is explicitly adjusted to avoid bad solutions. All these features make SPC different from the majority of the existing penalization frameworks, which either encourage sparsity in cluster means or use a convex penalty for sparsity. 

There are several questions that need to be addressed by future research. First, we have not discussed what can be considered a cluster and what can be considered noise. If a cluster consists of just one observation, it could safely be assumed that this singleton cluster is an outlier, however, a decision needs to be made, based on the data, with respect to clusters of size $N_k > 1$. This is easier to determine with smaller datasets, where the cut off $N_k > 1$ is usually sufficient, however, it will not be clear with larger and more complex data that might need higher cluster size cutoff values. We would like to address this issue in the future in a more principled manner. In this paper, we also only cursorily discuss the methods for selecting the best candidate solution from a solution path and we do not address the asymptotic properties of the cluster center estimates. These and other algorithmic improvements are interesting topics for future work.

\appendix
\section*{Appendix}

\subsection{Soft thresholding} 

In order to split an existing cluster, we cycle through every $\theta_i$ for $i =1, \ldots, n$ while fixing $\theta_{[-i]} = ( \theta_1, \ldots , \theta_{i-1}, \theta_{i+1}, \ldots, \theta_{n} )$ to its current value and minimize 
\begin{align}
	\label{eq:sthloss}
	& \ell( \theta_i ) = \| y_i - \theta_i \|_2^2 + \lambda \sum_{ k=1 }^{K} N_{k} \rho \left( \| \theta_i - \mu_{k} \|_2 \right),
\end{align}
over $\theta_i$. Suppose that $\theta_i^{(t)} = \mu_{k}$ for some $k$, i.e. the object $y_i$ has been assigned to the $k$th existing cluster. We cycle through each component of $\theta_i = ( \theta_{i1}, \ldots, \theta_{ip} )$ and write the difference in each component as $\beta_m = \theta_{im} - \mu_{km} $, $m=1,\ldots, p$. Let
\begin{align}
	\label{eq:sththeta}
	& \tilde{\theta}_{im}^{(t+1)} = \frac{ y_{im} + \lambda \sum_{ \ell \neq k } w_{i,\ell}^{(t)}  \mu_{\ell m} }{  1 + \lambda \sum_{ \ell \neq k } w_{i,\ell}^{(t)} },
\end{align}
and
\begin{align*}
	& \gamma= \frac{ \lambda N_{k}} { 1 + \sum_{\ell \ne k} \frac{\lambda N_{\ell}} { 2 \| \theta_{i}^{(t)} - \mu_{\ell} \|_2 }  }.
\end{align*} 
Minimization of $\ell(\theta_i)$ in \eqref{eq:sthloss} reduces to soft thresholding of $\hat{\beta}_m = \tilde\theta_{im}^{(t+1)} - \mu_{km}$. Let
\begin{align*}
	& \beta_m^{(t+1)}
	= \left\{
	\begin{array}{lll}
	\hat{\beta}_m - \gamma , & \hat{\beta}_m > 0 \text{ and } \gamma < | \hat{\beta}_m | \\
	\hat{\beta}_m + \gamma, & \hat{\beta}_m < 0 \text{ and } \gamma < | \hat{\beta}_m | \\
	0, & \gamma \geq | \hat{\beta}_m |.
	\end{array} \right.
\end{align*}
If $\beta^{(t+1)}_m = 0$ for all $m$, then $\theta_{i}^{(t+1)} = \theta_i^{(t)} = \mu_k$ and the corresponding $y_i$ will stay in its current cluster $k$. If $\beta^{(t+1)}_m \neq 0$ for any $m$, then the corresponding $y_i$ will be separated or unfused from its current cluster. In this case, $\theta_i^{(t+1)} \neq \mu_{k}$ and the update for the remainder of the coordinates $\left( \theta_{i(m+1)}, \ldots, \theta_{i\left(p\right)} \right)$ will be performed jointly as for $\theta_{im}^{(t+1)} $ \eqref{eq:sththeta}. Algorithm~\ref{alg:MM} could be modified accordingly to incorporate this split step.


\subsection{Proof of Lemma \ref{lemma1}} Fix $\theta_2^{(t)}$ and one minimization step of the MM algorithm as in \eqref{eq:mu} gives
\begin{align*}
&  \theta_1^{(t+1)} - \theta_2^{(t)}  = \frac{ y_1 + \lambda w_{1,2}^{(t)} \theta_2^{(t)} }{ 1 + \lambda w_{1,2}^{(t)} }  - \theta_2^{(t)}.
\end{align*} 
Equating this to $(1 -\phi)( \theta_1^{(t)} - \theta_2^{(t)} )$ and simply re-arranging the terms, we get
\begin{align*}
& \lambda = \frac{ 2 \eta \| ( \theta_1^{(t)} - y_1 ) + \phi ( \theta_2^{(t)} - \theta_1^{(t)} ) \|_2  }{ \left( \eta - \| \theta_1^{(t)} - \theta_2^{(t)} \|_2 \right)_+ (1 - \phi ) }.
\end{align*} 
For the initial step of the MM algorithm $\theta_1^{(0)} = y_1$ and $\theta_2^{(0)} = y_2$, and thus
\begin{align*}
& \lambda = \frac{  2 \phi \eta \| y_1 - y_2 \|_2 }{ (1 - \phi ) \left( \eta -  \| y_1 - y_2 \|_2 \right) },
\end{align*}
if $\eta > d = \| y_1 - y_2 \|_2$. If $\eta \leq d$, then $w_{1,2}^{(t-1)}=0$ and thus $\theta_i^{(t)}=y_i$ for all $t\geq 1$.

\subsection{Proof of Lemma \ref{lemma2}} 
Let $\gamma = \theta_2 - \theta_1$. Recall that $\ell( \theta_1, \theta_2)$ is minimized at 
$\left( \bar{y} - \gamma/2, \bar{y} + \gamma/2 \right)$ for any $\gamma$ and thus, 
minimizing \eqref{eq:twoloss} reduces to minimizing $\ell(\gamma)$ defined in \eqref{eq:twoloss_gamma}.
First assume that $\gamma=c(y_2-y_1)$ for $c\in \mathbb{R}$ and define
\begin{equation*}
f(c)\defi\ell(c(y_2-y_1))=\frac{1}{2} d^2 (c-1)^2 + \lambda \rho(d \cdot |c|),
\end{equation*}
where $d=\|y_2-y_1\|_2$. It is easy to see that $f(c)>f(0)$ for all $c<0$ and $f(c)>f(1)$ for all $c>1$.
Therefore, $f$ must be minimized globally at $c\in[0,1]$.
Then a sufficient condition for
$f(c)$ to be minimized at $c=0$, which gives $\gamma=0$, is 
\begin{align*}
\frac{\partial f (c)}{ \partial c } = d^2 (c-1) + \left( \lambda  - \frac{ d c }{ \delta } \right)_{+} d > 0,
\end{align*}
for $0<c\leq 1$.
If \eqref{eq:lemma2} holds, the above inequality is equivalent to
\begin{align*}
 \lambda  > \left[(1-c)+ \frac{  c }{ \delta }\right] d,
\end{align*}
which is obviously implied by \eqref{eq:lemma2}, noting that $c\in(0,1]$.
Lastly, for any nonzero $\gamma'$ that is not collinear with $y_2-y_1$, there is a $\gamma=c(y_2-y_1)$
such that $\|\gamma\|_2=\|\gamma'\|_2$ and $\|\gamma-(y_2-y_1)\|_2<\|\gamma'-(y_2-y_1)\|_2$
by triangle inequality, and thus $\ell(\gamma)<\ell(\gamma')$. 
Therefore, the global minimizer of \eqref{eq:twoloss} is 
$(\hat{\theta}_1,\hat{\theta}_2)=(\bar{y} ,\bar{y} )$ if \eqref{eq:lemma2} holds.

\subsection{ARI} 

Let $h(k) = \left( \begin{array}{c} k \\ 2 \end{array} \right)$. Given a contingency table $(n_{ij} )_{I \times J}$ with entries $n_{ij}$, row sums $n_{i \bullet} = \sum_{j} n_{ij}$, column sums $n_{\bullet j } = \sum_{i} n_{ij}$, and a total sum of entries $n = \sum_{i,j} n_{ij}$, the ARI is defined by
\begin{align}
\label{eq:ari}
& \text{ARI} = \frac{ \displaystyle\sum_{i=1}^{I} \displaystyle\sum_{j=1}^{J} h \left( n_{ij} \right) -  \left[ \displaystyle\sum_{i=1}^{I} h \left( n_{i \bullet} \right) \displaystyle\sum_{j=1}^{J} h\left( n_{\bullet j} \right) \right]    /   h \left( n \right) } { \frac{1}{2}  \left[  \displaystyle\sum_{i=1}^{I} h \left(  n_{i \bullet} \right) + \displaystyle\sum_{j=1}^{J} h \left( n_{ \bullet j } \right) \right] -  \left[  \displaystyle\sum_{i=1}^{I} h \left( n_{i \bullet} \right) \displaystyle\sum_{j=1}^{J} h \left( n_{\bullet j } \right) \right] / h \left( n \right) }.
\end{align}


\begin{thebibliography}{9}

\bibitem[Aggarwal and Reddy(2013)]{aggarwal2013} 
Aggarwal, C.C., Reddy, C.K.
(2013).
\emph{Data Clustering: Algorithms and Applications.}
Chapman \& Hall/CRC Data Mining and Knowledge Discovery Series,
Chapman and Hall/CRC.

\bibitem[Arthur and Vassilvitskii(2007)]{arthur2007} 
Arthur, D., Vassilvitskii, S.
(2007).
K-means++: The advantages of careful seeding.
In \emph{Proceedings of the Eighteenth Annual ACM-SIAM Symposium on Discrete Algorithms},
SODA '07,
1027-1035,
Philadenphia, PA, USA, 2007. Society for Industrial and Applied Mathematics.


\bibitem[Ben-Hur \textit{et al.}(2001)]{benhur2001} 
Ben-Hur, A., Horn, D., Siegelmann, H.T., Vapnik, V.
(2001).
Support vector clustering.
\emph{J. Mach. Learn. Res.}
\textbf{2} 125-137.


\bibitem[Bensmail \textit{et al.}(1997)]{bensmail1997} 
Bensmail, H., Celeux, G., Raftery, A.E., Robert, C.
(1997).
Inference in model-based cluster analysis.
\emph{Statist. Comput.}
\textbf{7} 1-10.


\bibitem[Byers and Raftery(1998)]{byers1998} 
Byers, S., Raftery, A.E. 
(1998).
Nearest-neighbor clutter removal for estimating features in spatial point processes.
\emph{J. Amer. Statist. Assoc.}
\textbf{93} 577-584.



\bibitem[Chi and Lange(2013)]{chi2013} 
Chi, E.C., Lange, K. 
(2013).
Splitting methods for convex clustering.
\emph{arXiv:1304.0499 [stat.ML]}.



\bibitem[De Smet \textit{et al.}(2002)]{desmet2002} 
De Smet, F., Mathys, J., Machal, K., Thijis, G., De Moor, B., Moreau, Y. 
(2002).
Adaptive quality-based clustering of gene expression profiles.
\emph{Bioinformatics}
\textbf{18} 735-746.


\bibitem[Efron \textit{et al.}(2004)]{efron2004} 
Efron, B., Hastie, T., Johnstone, I., Tibshirani, R.  
(2004).
Least angle regression (with discussion).
\emph{Ann. Statist.}
\textbf{32} 407-499.


\bibitem[Fan and Li(2001)]{fan2001} 
Fan, J., Li, R. 
(2001).
Variable selection via non-concave penalized likelihood and its oracle properties.
\emph{J. Amer. Statist. Assoc.}
\textbf{96} 1348 - 1360.


\bibitem[Fang and Wang(2012)]{fang2012} 
Fang, Y., Wang, J. 
(2012).
Selection of the number of clusters via the bootstrap method.
\emph{Comput. Stat. Data Anal.}
\textbf{56} 468-477.


\bibitem[Forero, Kekatos, and Giannakis(2012)]{forero2012} 
Forero, P., Kekatos, V,. Giannakis, G.B.
(2012).
Robust clustering using outlier-sparsity regularization.
\emph{IEEE Transactions on Signal Processing}
\textbf{60} 4163-4177.


\bibitem[Fraley and Raftery(2002)]{fraley2002} 
Fraley, C., Raftery, A.E. 
(2002).
Model-based clustering, discriminant analysis, and density estimation.
\emph{J. Amer. Statist. Assoc.}
\textbf{97} 611-631.




\bibitem[Friedman \textit{et al.}(2007)]{friedman2007} 
Friedman, J., Hastie, T., Hofling, H., Tibshirani, R.  
(2007).
Pathwise coordinate optimization.
\emph{Ann. Appl. Statist.}
\textbf{1} 302-332.


\bibitem[Friedman, Hastie, and Tibshirani(2010)]{friedman2010} 
Friedman, J., Hastie, T., Tibshirani, R.  
(2010).
Regularization paths for generalized linear models via coordinate descent.
\emph{J. Statist. Softw.}
\textbf{33} 1-22.


\bibitem[Fu and Zhou(2013)]{fu2013} 
Fu, F., Zhou, Q.  
(2013).
Learning sparse causal Gaussian networks with experimental intervention: regularization and coordinate descent.
\emph{JASA}
\textbf{108} 288-300.



\bibitem[Garcia-Escudero \textit{et al.}(2008)]{garcia2008} 
Garcia-Escudero, L.A., Gordaliza, A., Matran, C., Mayo-Iscar, A.  
(2008).
A general trimming approach to robust cluster analysis.
\emph{Ann. Statist.}
\textbf{36} 1324-1345.


\bibitem[Guo \textit{et al.}(2010)]{guo2010} 
Guo, J., Levina, E., Michailidis, G., Zhu, J.  
(2010).
Pairwise variable selection for high-dimensional model-based clustering.
\emph{Biometrics}
\textit{66} 793-804.


\bibitem[Hastie \textit{et al.}(2000)]{hastie2000} 
Hastie, T., Tibshirani, R., Eisen, M.B., Alizadeh, A., Levy, R., Staudt, L., Chan, W., Botstein, D., Brown, P.
(2000).
Gene shaving as a method for identifying distinct sets of genes with similar expression patterns.
\emph{Genome Biol.}
\textbf{1}(2) 0003.1-0003.21.


\bibitem[Hastie, Tibshirani, and Friedman(2009)]{hastie2009} 
Hastie, T., Tibshirani, R., Friedman, J. 
(2009).
\emph{The Elements of Statistical Learning: Data Mining, Inference, and Prediction.}
Springer; 2nd ed, 
p. 463.


\bibitem[Hocking \textit{et al.}(2011)]{hocking2011} 
Hocking, T.D., Joulin, A., Bach, F., Vert, J-P.
(2011).
Clusterpath: an algorithm for clustering using convex fusion penalties.
\emph{28th international conference on machine learning},
2011.


\bibitem[Hubert and Arabie(1985)]{hubert1985} 
Hubert, J., Arabie, P.
(1985).
Comparing partitions.
\emph{J. Classif.}
\textbf{2} 193-218.


\bibitem[Jain(2010)]{jain2010} 
Jain, A.K.
(2010).
Data clustering: 50 years beyond k-means.
\emph{Pattern Recognition Lett.}
\textbf{31}(8) 651-666.




\bibitem[Kohonen(1990)]{kohonen1990} 
Kohonen, T.
(1990).
The self-organizing map.
\emph{Proceedings of the IEEE}
\textbf{78} 1464-1479.


\bibitem[Kulis and Michael(2011)]{kulis2011} 
Kulis, B., Michael I.J. 
(2011).
Revisiting k-means: new algorithms via bayesian nonparametrics.
\emph{arXiv preprint arXiv:1111.0352}.


\bibitem[Lange(1995)]{lange1995} 
Lange, K.
(1995).
A gradient algorithm locally equivalent to the EM algorithm.
\emph{J. R. Statist. Soc. B}
\textbf{57} 425-437.



\bibitem[Lange, Hunter, and Yang(2000)]{lange2000} 
Lange, K., Hunter D.R., and Yang, I.
(2000).
Optimization transfer using surrogate objective functions.
\emph{J. Comput. Graph. Statist.}.
\textbf{9} 1-20.


\bibitem[Lange(2004)]{lange2004} 
Lange, K.  
(2004).
\emph{Optimization.}
Springer, New York,
119 - 133.


\bibitem[Lindsten, Ohlsson, and Ljung(2011)]{lindsten2011} 
Lindsten, F., Ohlsson, H., Ljung, L.
(2011).
Just relax and come clustering! a convexication of k-means clustering. 
\emph{Technical report from Automatic Control at Link{\"o}pings universitet},
2011.


\bibitem[Maitra and Ramler(2009)]{maitra2009} 
Maitra, R., Ramler, I.P. 
(2009).
Clustering in the presence of scatter.
\emph{Biometrics}
\textbf{65} 341-352.


\bibitem[Mazumder, Friedman, and Hastie(2011)]{mazumder2011} 
Mazumder, R., Friedman, J., Hastie, T.  
(2011).
SparseNet: coordinate descent with non-convex penalties.
\emph{J. Amer. Statist. Assoc.}
\textbf{106} 1125-1138.


\bibitem[McLachlan, Bean, and Peel(2002)]{mclachlan2002} 
Mclachlan, G.J., Bean, R.W., Peel, D. 
(2002).
A mixture model-based approach to the clustering of microarray expression data.
\emph{Bioinformatics}
\textbf{18} 413-422.




\bibitem[Oh and Raftery(2007)]{oh2007} 
Oh, M.S., Raftery, A.E. 
(2007).
Model-based clustering with dissimilarities: a bayesian approach.
\emph{J. Comput. Graph. Statist.}
\textbf{16} 559-585.


\bibitem[Pan and Shen(2007)]{pan2007} 
Pan, W., Shen, X.  
(2007).
Penalized model-based clustering with application to variable selection.
\emph{J. Mach. Learn. Res.}
\textbf{8} 1145-1164.


\bibitem[Pan, Shen, and Liu(2013)]{pan2013} 
Pan, W., Shen, X., Liu, B.  
(2013).
Cluster analysis: unsupervized learning via supervized learning with a non-convex penalty.
\emph{J. Mach. Learn. Res.}
\textbf{14} 1865-1889.


\bibitem[Pelckmans \textit{et al.}(2005)]{pelckmans2005} 
Pelckmans, K., De Brabanter, J., Suykens, J., De Moor, B.   
(2005).
Convex clustering shrinkage.
\emph{PASCAL Workshop on Statistics and Optimization of Clustering Workshop},
2005.


\bibitem[Sharan, Maron-Katz, and Shamir(2003)]{sharan2003} 
Sharan, R., Maron-Katz, A., Shamir, R.
(2003).
CLICK and EXPANDER: a system for clustering and visualizing gene expression data.
\emph{Bioinformatics}
\textbf{19} 1787-1799.


\bibitem[Shen, Sun, and Li(2010)]{shen2010} 
Shen, Y., Sun, W., Li, K,C.
(2010).
Dynamically weighted clustering with noise set.
\emph{Bioinformatics}
\textbf{26} 341-347.


\bibitem[Soltanolkotabi, Elhamifar, and Candes(2013)]{soltanolkotabi2013} 
Soltanolkotabi, M., Elhamifar, E., Candes, E.J.
(2013).
Robust subspace clustering. 
\emph{arXiv:1301.2603}.


\bibitem[Sun and Wang(2012)]{sun2012} 
Sun, W., Wang, J.
(2012).
Regularized k-means clustering of high-dimensional data and its asymptotic consistency. 
\emph{Elec. J. Stat.}
\textbf{6} 148-167.


\bibitem[Thalamuthu \textit{et al.}(2006)]{thalamuthu2006} 
Thalamuthu, A., Mukhopadhyay, I., Zheng, X., Tseng, G.C. 
(2006).
Evaluation and comparison of gene clustering methods in microarray analysis.
\emph{Bioinformatics}
\textbf{22} 2405-2412.


\bibitem[Tibshirani(1996)]{tibshirani1996} 
Tibshirani, R.
(1996).
Regression shrinkage and selection via the lasso.
\emph{JRSSB}
\textbf{58} 267-288.


\bibitem[Tibshirani \textit{et al.}(2005)]{tibshirani2005} 
Tibshirani, R., Saunders, M., Rosset, S., Zhu, J., Knight, K. 
(2005).
Sparsity and smoothness via the fused lasso.
\emph{J. Roy. Statist. Soc. Ser. B}
\textbf{67} 91-108.


\bibitem[Tibshirani and Walther(2005)]{tibshirani2005_1} 
Tibshirani, R., Walther, G.
(2005).
Cluster validation by prediction strength.
\emph{J. Comput. Graph. Statist.}
\textbf{14} 511-528.


\bibitem[Tibshirani, Walther, and Hastie(2001)]{tibshirani2001} 
Tibshirani, R., Walther, G., Hastie, T.
(2001).
Estimating the number of clusters in a data set via the gap statistic.
\emph{JJ. Roy. Statist. Soc. Ser. B}
\textbf{63} 411-423.




\bibitem[Tritchler, Parkhomenko, and Beyene(2009)]{tritchler2009} 
Tritchler, D., Parkhomenko, E., Beyene, J.
(2009).
Filtering genes for cluster and network analysis.
\emph{BMC Bioinformatics}
\textbf{10}:193.


\bibitem[Tseng and Wong(2005)]{tseng2005} 
Tseng, G.C., Wong, W.H.
(2005).
Tight clustering: a resampling-based approach for identifying stable and tight patterns in data.
\emph{Biometrics}
\textbf{61} 10-16.


\bibitem[Tseng(2007)]{tseng2007} 
Tseng, G.C. 
(2007).
Penalized and weighted k-means for clustering with scattered objects and prior information in high-throughput biological data.
\emph{Bioinformatics}
\textbf{23} 2247-2255.


\bibitem[Tseng and Yun(2009)]{tseng2009} 
Tseng, P., Yun, S.
(2009).
A coordinate gradient descent method for nonsmooth separable minimization.
\emph{Mathematical Programming}
\textbf{117} 387-423.


\bibitem[von Luxburg(2007)]{luxburg2007} 
von Luxburg, U. 
(2007).
A tutorial on spectral clustering.
\emph{Stat. Comput.}
\textbf{17} 395-416.


\bibitem[Wang, Neill, and Miller(2008)]{wangH2008} 
Wang, H., Neill, J., Miller, F.  
(2008).
Nonparametric clustering of functional data.
\emph{Stat. Interface}
\textbf{1} 47-62.


\bibitem[Wang and Zhu(2008)]{wang2008} 
Wang, S., Zhu, J.  
(2008).
Variable selection for model-based high-dimensional clustering and its applications to microarray data.
\emph{Biometrics}
\textbf{64} 440-448.


\bibitem[Witten and Tibshirani(2010)]{witten2010} 
Witten, D., Tibshirani, R.
(2010).
A framework for feature selection in clustering.
\emph{J. Amer. Statist. Assoc.}
\textbf{105} 713-726.


\bibitem[Wu and Lange(2008)]{wu2008} 
Wu, T.T., Lange, K.  
(2008).
Coordinate descent algorithms for lasso penalized regression.
\emph{Ann. Appl. Statist.}
\textbf{2} 224-244.


\bibitem[Xie, Pan, and Shen(2008)]{xie2008} 
Xie, B., Pan, W., Shen, X.  
(2008).
Variable selection in penalized model-based clustering via regularization on grouped parameters.
\emph{Biometrics}
\textbf{64} 921-930.


\bibitem[Yeung \textit{et al.}(2001)]{yeung2001} 
Yeung, K.Y., Fraley, C., Murua, A., Raftery, A.E., Ruzzo, W.L. 
(2001).
Model-based clustering and data transformations for gene expression data.
\emph{Bioinformatics}
\textbf{17} 977-987.


\bibitem[Yuan and Lin(2006)]{yuan2006} 
Yuan, M., Lin, Y.  
(2006).
Model selection and estimation in regression with grouped variables.
\emph{J. Roy. Statist. Soc. Ser. B}
\textbf{68} 49-67.




\bibitem[Zhang(2010)]{zhang2010} 
Zhang, C.H. 
(2010).
Nearly unbiased variable selection under minimax concave penalty.
\emph{Ann. Statist.}
\textbf{38} 894-942.






\bibitem[Zhou, Pan, and Shen(2009)]{zhou2009} 
Zhou, H., Pan, W., Shen, X.
(2009).
Penalized model-based clustering with unconstrained covariance matrices.
\emph{Elec. J. Stat.}
\textbf{3} 1473-1496.


\bibitem[Zhou \textit{et al.}(2007)]{zhou2007} 
Zhou, Q., Chipperfield, H., Melton, D.A., Wong, W.H. 
(2007).
A gene regulatory network in mouse embryonic stem cells.
\emph{PNAS}
\textbf{104} 16438-16443.


\bibitem[Zou and Hastie(2005)]{zou2005} 
Zou, H., Hastie, T. 
(2005).
Regularization and variable selection via the elastic net.
\emph{J. Roy. Statist. Soc. Ser. B}
\textbf{67} 301-320.







\end{thebibliography}
\end{document}